\documentclass[12pt]{iopart}
\usepackage{fullpage}
\usepackage{threeparttable}
\usepackage{pdflscape}
\usepackage{longtable}

\usepackage[position=b,singlelinecheck=off]{subfig}  %% for using subfig or subfloat
\usepackage{psfrag}
\usepackage{graphicx}
\usepackage{epstopdf}
\usepackage{color}
\usepackage{lineno}
\usepackage{hyperref}
\usepackage[numbers,sort&compress]{natbib}
\usepackage{multirow}
\usepackage{multicol}
\usepackage{tikz}
\usepackage{refcount}
\usepackage{tensor}
\usepackage{adjustbox}
\usepackage{tablefootnote}

\newcommand{\inprog}[1]{{\textcolor{blue}{\tt In Progress}}}

\newcommand{\Sf}{$S_{\rm f}$}%
\newcommand{\Af}{$A_{\rm f}$}%
\newcommand{\ionF}{$F_{\rm ion}$}%
\newcommand{\Un}{${|{\fU}| }^2$}%
\newcommand{\Ta}{T$=0.6\,T_{\rm m}$}%
\newcommand{\Tb}{T$=0.7\,T_{\rm m}$}%
\newcommand{\Tc}{T$=0.8\,T_{\rm m}$}%
\newcommand{\Ra}{$R=2\,$nm}%
\newcommand{\Rb}{$R=4\,$nm}%
\newcommand{\Rc}{$R=6\,$nm}%
\newcommand{\potB}{BMH}%
\newcommand{\potV}{\textsc{Vash}}%
\newcommand{\potC}{CB}%
\newcommand{\potCT}{CTIE}%
\newcommand{\ans}[1]{$\mathring{\rm A}$}
\newcommand{\anss}[1]{$\mathring{\rm A}^2$}
\newcommand{\ith}[1]{$i^{\rm th}$}
\newcommand{\jth}[1]{$j^{\rm th}$}
\newcommand{\ovito}{\textsc{Ovito}}

\def\etal{\emph{et al.}~}
\newcommand{\fr}{\bi{r}}
\newcommand{\fF}{\bi{F}}
\newcommand{\fU}{\bi{U}}
\newcommand{\fR}{\bi{R}}

\begin{document}

%\begin{frontmatter}
%\title{Sintering of Alumina Nano Particles with Molecular Dynamics Simulation}
\title{Sintering of Alumina Nanoparticles: Comparison of Interatomic Potentials, Molecular Dynamics Simulations, and Data Analysis}%\tnoteref{mytitlenote}}

\author{S. Roy$^{1,2}$\footnote[1]{These authors contributed equally.}, 
        A. Prakash$^{2}$ \footnote[1]{These authors contributed equally.},
        and S. Sandfeld$^{1, 2, 3, 4}$%
    }

\address{$^1$%
	Institute for Advanced Simulation: 
	Materials Data Science and Informatics (IAS-9),
	Forschungszentrum Juelich GmbH, 52425 Juelich, Germany%
}
\address{$^2$%
	Chair of Micromechanical Materials Modelling (MiMM), 
	Institute of Mechanics and Fluid Dynamics, 
	TU Bergakademie Freiberg, 09599 Freiberg, Germany%
}
\address{$^3$%
    JARA, Forschungszentrum Juelich GmbH, 52425 Juelich, Germany%
}
\address{$^4$%
    Chair of Materials Data Science and Materials Informatics
    Faculty 5, RWTH Aachen University, 52056 Aachen, Germany
}

\ead{s.sandfeld@fz-juelich.de}

\begin{abstract}
Sintering of alumina nanoparticles is of interest both from the view of fundamental research as well as for industrial applications. Atomistic simulations are tailor-made for understanding and predicting the time- and temperature-dependent sintering behaviour. However, the quality and predictability of such analysis is strongly dependent on the performance of the underlying interatomic potentials. 
In this work, we investigate and benchmark four empirical interatomic potentials and discuss the resulting properties and drawbacks based on experimental and density functional theory data from the literature. The potentials, which have different origins and formulations, are then used in molecular dynamics simulations to perform a systematic study of the sintering process. To analyse the results, we develop a number of tailored data analysis approaches that are able to characterise and quantify the sintering process. Subsequently, the disparities in the sintering behaviour predicted by the potentials are critically discussed. Finally, we conclude by providing explanations for the differences in performance of the potentials, together with recommendations for molecular dynamics sintering simulations of alumina.
\end{abstract}

%\begin{keyword}
%alumina, interatomic potential, molecular dynamics simulations, sintering 
%\end{keyword}
%\keywords{alumina, interatomic potential, molecular dynamics simulations, sintering}
%\keywords{magnetic moment, solar neutrinos, astrophysics}
%\end{frontmatter}

\submitto{\MSMSE}

%%%%%%% INTRODUCTION %%%%%%%%%%%%%%%%%%%%%%%%%%%
\section{Introduction}\label{sec:intro}
 
Aluminium oxide (Al$_2$O$_3$) or \emph{alumina} has evoked significant research interest due to its use in technological and industrial  applications, ranging from the field of aerospace~\cite{rathod:2017}, water treatment, filtration and cooling systems~\cite{bhatnagar:2010, sarkar:2012, sridhara:2011}, medical and health care~\cite{sadiq:2009, becker:2016,mahmoudian:2019}, to materials engineering~\cite{chevalier:2009, hulbert:1993, benzaid:2008}. Alumina coating on aluminium is known to improve wear resistance~\cite{valant:2016} and act as electrical and thermal insulation~\cite{he:2009}. The material exhibits unique properties, such as e.g., high thermal conductivity, high hardness, and a good resistance to corrosion and abrasion \cite{kasprzyk:2004}. 

An important technique for processing alumina is sintering, which involves consolidation and densification of powder compacts, and is driven by a reduction in total interface energy \cite{kang:2004}. As one of the technologies that has been known and used for a long time, the process has proven to be important and beneficial for the production of a variety of ceramic structures, from powder-metallurgy parts to bulk components. Recently, sintering has evoked much interest in modern industrial production technologies such as 3D printing and additive manufacturing. 

Accurate modelling of the sintering process is on the wish list of many sintering industries, mainly due to the engineering expediency it provides. However, the complex interplay of a multitude of parameters together with a lack of fundamental materials understanding at the nanoscale have proven to be a major hindrance to this end 
\cite{german:2002}.

Indeed, many of the benefits and applications of alumina have their origins at the nanoscale making the material system and the sintering process ideally suited for investigation with atomistic modelling and simulations. Such simulations, which follow the trajectory of individual atoms, have now established themselves as an indispensable tool in advancing our understanding of materials behavior~\cite{Farkas:2013,Wunderlich:2014,Prakash:2017}. Sintering and concomitant processes have been investigated through molecular dynamics (MD) simulations for a large variety of materials (including alumina) and for a number of different techniques such as selective laser sintering or  pressure-assisted/pressure-less sintering~\cite{nandy:2019, song:2010, koparde:2005}. 

Atomistic modelling of materials behavior is, however, not straightforward and hinges primarily on three aspects~\cite{Prakash:2017}: i) accuracy of the initial structure used, ii) reflection of real-world boundary conditions, and most importantly, iii) the reliability, robustness and predictive capability of the interatomic potential used. Such interatomic potentials are particularly critical for oxide materials such as alumina due to the ionic (and partly covalent) nature of the oxygen-metal bonds, which require the inclusion of long-range effects. Although many interatomic potentials for alumina exist, each potential is developed with a different rationale in terms of applicability and transferability. The choice of the interatomic potential hence becomes crucial for the atomistic study, and is usually made by evaluating the predictive capability of the potential on certain basic material properties -- e.g., lattice constants, cohesive and surface energies and elastic constants -- and subsequently applied to the more complex problem at hand.

Aluminium oxide often exists in crystalline form -- the \emph{corundum structure} -- and is a member of the hexagonal close packed (hcp) family. A single molecule consists of two aluminium and three oxygen atoms, the interactions among which are usually modelled by incorporating two-body and/or three-body interaction terms. The Coulomb-Buckingham~\cite{houska:2013} or Born-Mayer-Huggins type potentials~\cite{bouhadja:2013} are based purely on two-body interactions, whilst the potential by Vashishta \emph{et al.}~\cite{vashishta:2008} incorporates both two-body and three-body interactions, with the latter associated with triplets containing Al-O bonds. A common feature of such potentials is the usage of fixed charges -- both nominal and effective -- for Al and O ions. To account for local heterogeneous electrostatic environment, particularly in the presence of interfaces, free surfaces and segregated ions, improvements have been proposed by incorporating variable point charges. Reactive potentials, such as reactive empirical bond order (REBO), reactive force field (ReaxFF) and charge optimised bond order  (COMB) potentials, integrate many-body effects and two-body interactions through a bond order term~\cite{liang:2013}. Other approaches such as the charge-transfer ionic~\cite{streitz:1994} and the second-moment tight-binding~\cite{rappe:1991} models are simpler in their approach and allow for charge equilibration during the process under study.

In this work, we investigate the sintering of spherical nanoparticles using molecular dynamics simulations. Specifically, we compare four different interatomic potentials -- the Vashishta potential~\cite{vashishta:2008}, the Coulomb-Buckingham potential of the Matsui type~\cite{matsui:1996}, the Born-Mayer-Huggins potential parametrised by Bouhadja \emph{et al.}~\cite{bouhadja:2013} and the charge transfer ionic potential~\cite{streitz:1994}. We first present the mathematical formulations of the potentials and evaluate the predictive capability of these potentials towards basic material properties. Subsequently, the sintering process is simulated for three different temperatures by considering spherical nanoparticles of three different sizes. Sintering is modelled as isothermal, high temperature process. In other words, we neglect pressure effects that play an important role in low temperature sintering~\cite{biesuz:2020, grass:2020, vakifahmetoglu:2020}. We develop a data analysis strategy that is based on a number of atomic and geometric properties to quantify and characterize the process. As a result, we find that despite exhibiting similar material properties, the investigated potentials result in very different sintering outcomes. We then discuss the characteristics used in terms of their ability to clearly quantify the progress of sintering by setting numerical thresholds. 

In the following section, we start by introducing the used potentials and rewrite them in a consistent mathematical formulation. Subsequently, details of modelling, simulation, and data analysis for characterising sintering are provided. We then compare various alumina properties (such as elastic properties, surface energies, lattice parameters, etc.) that result from our atomistic simulations with data from experiments and density functional theory calculations. The actual sintering simulations are analysed in detail for all four potentials, which is concluded by a discussion of their respective suitability and appropriateness for MD simulations of nanoparticle sintering of alumina.

\section{Interatomic potentials of alumina}\label{sec:potentials}
 
In the following, four interatomic potentials are considered for the sintering study: the Vashishta potential (in the following abbreviated as {\potV}), the Coulomb-Buckingham potential ({\potC}), Born-Mayer-Huggins potential ({\potB}), and the Charge Transfer Ionic + EAM potential ({\potCT}). Whilst the last three are based on two-body interactions, the first one, i.e. \potV{} is based on both two-body and three-body interactions. Three potentials ({\potV}, {\potB}, and {\potC}), have the same fixed charges for the anions and cations, and one ({\potCT}), allows for variable charges. In the following, all the potentials used in our study are briefly introduced.
 %
% Three of them are based on two-body interactions, the fourth ({\potV}) is based on two- and three-body interactions. 

 \subsection{Three-body + two-body interaction potential: the Vashishta potential}
% \subsubsection{The Vashishta potential}   
 %The Vashishta potential 
Developed by Vashishta~\etal~\cite{vashishta:2008}, \potV{} is an interatomic potential based on three-body interaction among Al and O atoms as O-Al-Al and Al-O-O as well as two-body interaction as Al-Al, O-O and Al-O. It reads
 \begin{equation} \label{eqn:vashishta08}
 U = \sum_{i<j} U^{(2)}_{ij} (r_{ij}) + \sum_{i,j<k} U^{(3)}_{jik} (r_{ij}, r_{ik}),
 \end{equation} 
   where the two-body part ($U^{(2)}_{ij} (r_{ij})$) of the effective potential is written as
 \begin{equation} \label{eqn:vashishta_2body}
 U^{(2)}_{ij} (r) = \frac{H_{ij}}{r^{\eta_{ij}}} + \frac{Z_i Z_j}{r_{ij}} {\rm exp}({-r_{ij}/\lambda}) - \frac{D_{ij}}{2r_{ij}^4} {\rm exp}(-r_{ij}/\zeta) - \frac{W_{ij}}{r_{ij}^6}.
 \end{equation} 
 Here, $H_{ij}$ is the strength of the steric repulsion, $Z_i$ the effective charge (in units of the electronic charge $|e|$), $D_{ij}$ is the strength of charge-dipole attraction, $W_{ij}$ is the van der Waals interaction strength, $\eta_{ij}$ the exponents of the steric repulsion term, $r_{ij}=|\fr_i-\fr_j|$ the distance between the $i^{\rm th}$ and $j^{\rm th}$ atoms, and $\lambda$ and $\zeta$ are the screening lengths for Coulomb and charge-dipole terms, respectively. The two-body interaction potential includes steric size effects of the ions, charge-transfer effects leading to Coulomb interactions, charge-dipole interactions due to the electronic polarisability of ions and induced dipole-dipole van der Waals interactions. It has a cut off at $r_c$ ($r_c=6$~\ans\; for alumina) beyond which the tow-body term is zero. 
 The three-body term is as follows:
 \begin{equation} \label{eqn:vashishta_3body}
 U^{(3)}_{jik} (r_{ij}, r_{ik}) = R^{(3)}(r_{ij}, r_{ik}) P^{(3)} (\theta_{jik}),    
 \end{equation} 
 where
 \begin{equation} \label{eqn:vashishta_3body1}
 R^{(3)}(r_{ij}) = B_{jik} \exp \Big(\frac{\gamma}{r_{ij}-r_0} + \frac{\gamma}{r_{ik}-r_0} \Big) \Theta (r_0-r_{ij}) \Theta(r_0-r_{ik}),
 \end{equation} 
 
 \begin{equation} \label{eqn:vashishta_3body2}
 P^{(3)} (\theta_{jik}) = \frac{(\cos \theta_{jik} - \cos \bar\theta_{jik})^2 } {1+ C_{jik} (\cos \theta_{jik} - \cos \bar \theta_{jik} )^2 }.
 \end{equation} 
 In Eq.~\ref{eqn:vashishta_3body}, \ref{eqn:vashishta_3body1} and \ref{eqn:vashishta_3body2}, $B_{jik}$ is the strength of the interaction, $\theta_{jik}$ the angle formed by $\fr_{ij}$ and $\fr_{ik}$,  $C_{jik}$ and $\bar \theta_{jik}$ are constants, and $\Theta (r_0- r_{ij})$ is a step function. The three-body term becomes zero when $\theta_{jik}=\bar \theta_{jik}$. %, a given constant for the angle.

 \subsection{Two-body interaction potentials}
 \subsubsection{The Coulomb-Buckingham potential}
 
 %The Coulomb-Buckingham potential 
 The ({\potC}) potential
 consists of of long range (Coulomb) and short range (Buckingham) potential terms as shown in Eq~\ref{eqn:coulomb_bucking}. % doi:10.1016/j.physb.2009.07.193.
 \begin{equation} \label{eqn:coulomb_bucking} 
 U_{ij}(r_{ij})  = U^{\rm Coulomb}_{\rm long} + U^{\rm Buckingham}_{\rm short} \\
 = \frac{1}{4 \pi \epsilon_0} \frac{Z_i Z_j}{r_{ij}} +  A_{ij} \exp \big(\frac{-r_{ij}}{\rho_{ij}} \big) - \frac{C_{ij}}{r_{ij}^{6}},
 \end{equation}
 where $Z_i$ and $Z_j$ are the effective charges,  $A_{ij}$ and $\rho_{ij}$ are the parameters of repulsion, $r_{ij}$\footnotemark 
 is the distance between the \ith\;~and \jth\;~atoms, $C_{ij}$ is the Van der Waals constant. 
 The exponential term in the {\potC} potential provides a better description of strong repulsion due to the overlap of the closed shell electron clouds, which is vital in a simulation of bombardment by energetic atoms or ions, etc. When atoms lie in a crystalline environment, the distance between two atoms is small, and the short-range potential plays an important role. However, when atoms are far from the substrate (free atoms), the larger distances between atoms leads to a rapid reduction of the short-range contribution and the long-range potential has the dominant effect on the free atoms. This potential can be categorised as~\cite{houska:2013} %http://dx.doi.org/10.1016/j.surfcoat.2013.07.062
 \begin{enumerate}
 	\item partial-charge $(Al^{+1.4175}, O^{-0.945})$ Buckingham-type Matsui~\cite{matsui:1996}
 	\item full-charge $(Al^{+3}, O^{-2})$ Buckingham-type Bacorisen potential~\cite{bacorisen:2006}
 	\item full-charge $(Al^{+3}, O^{-2})$ Buckingham-type Sun potential~\cite{sun:2006}
	\item BKS (Beest-Krammer-Santen) potential; which is another {\potC} type potential~\cite{hoang:2004}. 
 \end{enumerate}
 In the current study we use as {\potC} potential the  Matsui type.

 \subsubsection{The Born-Mayer-Huggins or Tosi/Fumi potential}
 
  The {\potB} potential consists of long range (Coulomb) and a short range interactions terms, %\todo{why do we have a cutoff for BMH? Is this for the Wolff summation? If yes, then this should be removed since the interactions are not cut off}:
  \begin{equation} \label{eqn:coulomb_bmh}
 U_{ij} = \frac{1}{4 \pi \epsilon_0} \frac{Z_i Z_j}{r_{ij}} + A \exp\Big(\frac{\sigma-r_{ij}}{\rho}\Big) - \frac{C}{r_{ij}^6} + \frac{D}{r_{ij}^8}, %\;\; \; r_{ij}<r_c,
 \end{equation}
 where $\sigma$ is an interaction-dependent length parameter, $\rho$ is an ionic-pair dependent length parameter. 
 The first term represents the long-range Coulomb interaction with charges $Z_i$ and $Z_j$ between $i^{\rm th}$ and $j^{\rm th}$ ions, separated by $r_{ij}$\footnotemark[\value{footnote}]
 \footnotetext{Implementation of  \potC{} and \potB{} potentials in LAMMPS requires cut-off distances for non-coulombic and coulombic  interactions, and $10$ and $15$ \ans{arg1} are used as cut-off values for those, respectively.}
  The last three terms on the right hand side represent the Born repulsive, van der Waals and  dipole dispersion interactions, respectively. %$r_c$ is the cut off distance \todo{for what?} of the potential. 
 In the current study, we use this potential parameterised by \citep{bouhadja:2013}.
 %The second term is the Born repulsive term which results from electronic overlap of neighboring atoms, while the two last contributions are the two first terms of the dipolar expan- sion, though here we take only the van der Waals term into account. 
 %The four terms on the right hand side represent the Coulombic, repulsive, van der Waals and  dipole dispersion interactions, respectively. In the current study, we use this potential parameterized by \citep{bouhadja:2013}.

\subsubsection{The Charge Transfer Ionic + EAM potential}

 The {\potCT} potential
 %The Charge Transfer Ionic + EAM ({\potCT}) potential 
 consists of contributions of the non-ionic interaction and those of the ionic interaction and charge transfer~\cite{streitz:1994, zhou:2004},
\begin{eqnarray}\label{eqn:cti_eam}
	U_{ij}	= U^{\rm CTI}_{ij}+U^{\rm EAM}_{ij}, 
\end{eqnarray}
where $U^{\rm CTI}$ is the {\potCT} potential for ionic interaction and charge transfer, %for a point charge of charge $q_i$, $V_{ij} = \frac{q_iq_j}{r_{ij}} $, 
and $U^{\rm EAM}$ is the non-ionic interaction.  The first part becomes zero for non-electrostatic material.  The electrostatic part can be expressed as:
\begin{equation}
	 U^{\rm CTI}_{ij} = U_0+\sum_{i} q_i \chi_i	 + \frac{1}{2}\sum_{i,j}q_i q_j V_{ij}, 
\end{equation}
with,
\begin{eqnarray}
	 	 \chi_i = \chi_0 + \sum_{j} Z_i  (\frac{j}{f_i} - \frac{f_i}{f_j}) ,  \\ %f(f_i, f_j) \\	 	 
	 V_{ij} = J^0_i \delta_{ij} + \sum_{r_i} \frac{f_i}{f_j}, \\
	 f_i = f_i(|r-r_i|) = \frac{\zeta^3}{\pi} {\rm exp} (-2\zeta r -r_i).	 	  
	  %f(f_i, f_j)	  
	  %	&= U_0 + \sum_i q_i \chi_i + \frac{1}{2}\sum_{i,j} q_iq_j V_{ij} + U_{EAM}(r) , %\\
%	&= U_0 + \sum_i q_i \chi_i + \frac{1}{2}\sum_{i,j} q_iq_j V_{ij} + F_{\alpha} \sum_{i\ne j} \rho_{\beta} (r_{ij}) + \frac{1}{2}\sum_{i\ne j} \phi_{\alpha \beta} (r_{ij}) 		
\end{eqnarray}
 The pair potential is written as 
 \begin{eqnarray}
 	U^{\rm EAM}_{ij} &=  \frac{1}{2} \sum_{i=1}^{N}  \sum_{j=i}^{i_N} \phi_{ij} (r_{ij}) + \sum_{i=1}^{N} F_i (\rho_i), 	         								
 \end{eqnarray}
where $\phi_{ij} (r_{ij})$ is the pair energy between the $i^{\rm th}$ and $j^{\rm th}$ atom separated by $r_{ij}$, $F_i({\rho_i})$ is the embedding energy to embed an atom $i$ in a local site with electron charge density $\rho_i$. More details are given in  \cite{streitz:1994, zhou:2004}.

\subsubsection{A first juxtaposition of the four potentials}
From the above equations we observe the following: The formulations of  {\potC} and {\potB} are mathematically similar and become identical if $D$ is zero in Eq.~\ref{eqn:coulomb_bmh}. They both are two-body interaction potentials. However, in the two-body interaction of the {\potV} potential, there are two additional terms for steric repulsion and charge-dipole interaction. The additional term for three-body interaction incorporated in {\potV} contains spatial and angular dependent factors, which is useful for amorphous alumina~\cite{vashishta:2008}. The three-body term applies to triplets Al-O-Al and O-Al-O with fixed angles. The charges are fixed and the same for anions and cations in all these potential. By contrast, {\potCT} is a potential of variable charges, where charges of two anions or cations are not identical.  Numerical values of the parameters used in the four potentials are tabulated in the \emph{Appendix}.

%%%%%%%%%%%% METHODS  %%%%%%%%%%%%%%%%%%%%%%%%
  \section{Methods} \label{sec:MD_method}
 All simulations are performed using the Large-scale Atomic/Molecular Massively Parallel Simulator (LAMMPS) atomistic simulation software~\cite{plimpton:1995}. Molecular statics (MS) is employed to calculate and verify basic potential properties of individual potentials, and molecular dynamics is employed to determine the melting temperature and to simulate sintering. A stable time step of 1~fs is used in all MD simulations.

\subsection{Material properties at 0~K}
Five different properties of the interatomic potentials, viz., lattice constants, cohesive energy, vacancy formation energies, elastic constants and surface energies, are calculated using standard procedures via MS simulations. Further details are provided in the\emph{supplementary material}.

\subsection{Melting temperature}
The melting temperature of alumina for the different potentials is determined using standard molecular dynamics simulations via the following procedure. A 3D triclinic sample of size $33 \times 29 \times 39$~\ans{arg1} and an $xy$ tilt factor of $16.7$~\ans{arg1}  %\todo{details about the sample (number of atoms, box size, etc.)} 
consisting of $1764$~Al and $2746$~O atoms and periodic boundary conditions in all directions is thermalised at several targeted temperatures using the  isobaric isothermal (NPT) ensemble. The average atomic volume, determined via a Voronoi tessellation as implemented in the open visualisation tool \ovito{}, is plotted for the different target temperatures. The melting point for the system and potential is then approximated as the temperature at which a jump in the average atomic volume is observed. For more details, the reader is referred to the \emph{supplementary material}.

\subsection{Sintering}
\begin{figure}
	\centering
  	\includegraphics[width=0.8\textwidth]{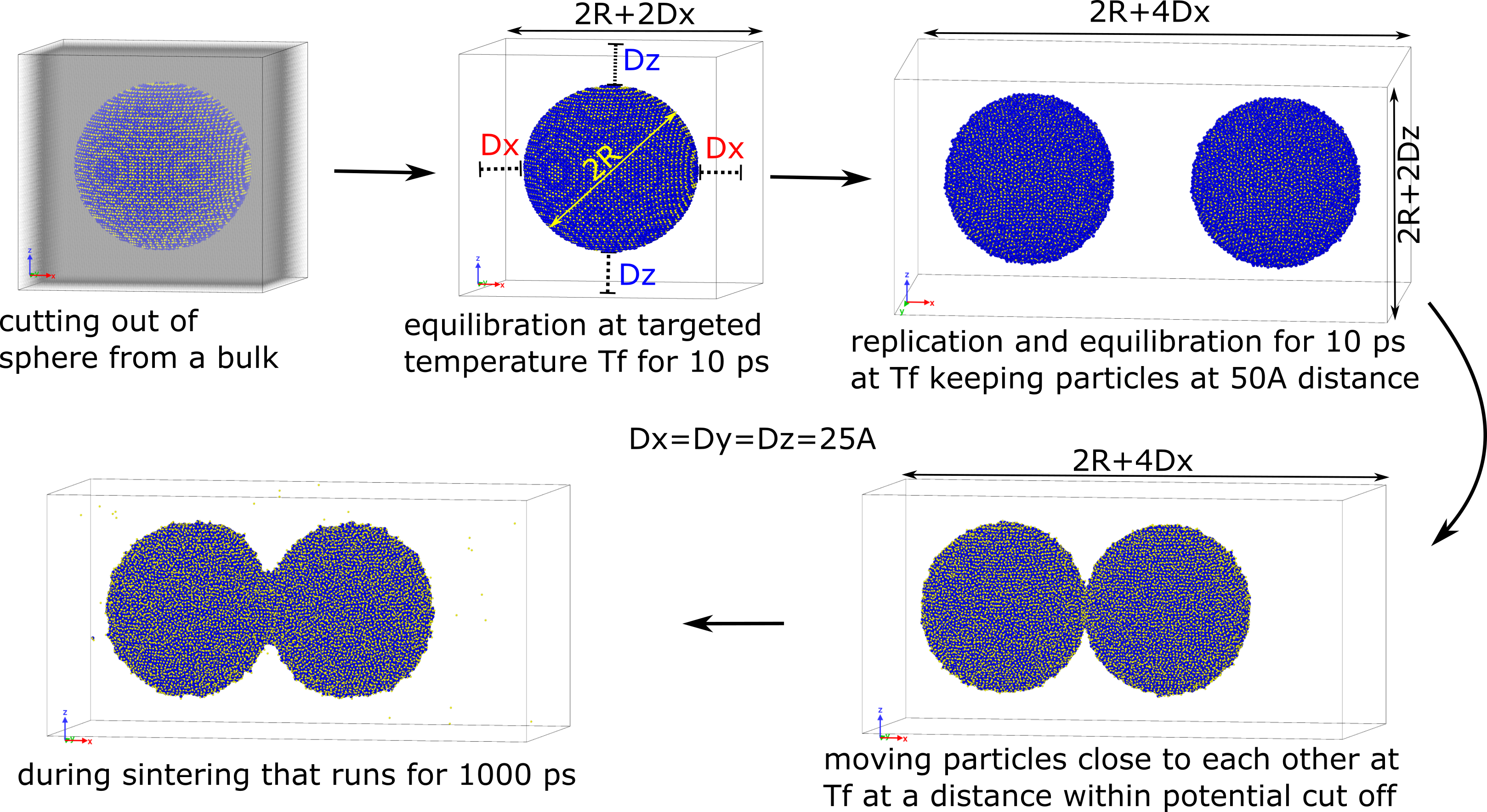}        
  	\caption{ The simulation setup: A particle is cut out from bulk material (shown as semi-transparent), where $D_x=D_y=D_z=25$~{\ans{A}} is the smallest distance between particle and the surfaces. Al and O atoms are colored with blue and yellow, respectively. The particle is heated up to the target temperature $T_{\rm f}$ and thermalised at $T_{\rm f}$ for $10$~ps each. It is then replicated and equilibrated at $T_f$. The particles are then brought together at a distance within the potential cut off and at $T_{\rm f}$. Subsequently, the system is maintained at the $T_{\rm f}$ for 1 ns during which sintering can occur under favorable conditions.}  	\label{fig:schematic_method}
\end{figure}

Fig.~\ref{fig:schematic_method} illustrates the approach used for simulating the sintering process.
Spherical single crystal nanoparticles (NPs) of three different radii $R=2$, $4$ and $6$~nm are considered for the solid-phase sintering simulations. A bulk $\rm Al_2O_3$ system, with $[11\bar 20]$, $[10 \bar10]$ and $[0001]$ in $x$, $y$ and $z$ directions  of the simulation box, respectively, is created using the atomic simulation environment tool~\cite{larsen:2017}. 
 %with the crystallographic $[1000] $ in $x$, direction \todo{will double check the orientation for y and z}. 
Spherical particles of different radii are cut out of the bulk system using Atomsk~\cite{hirel:2015}. 
%The crystallographic orientation of the spherical particles are $[11\bar 20]$, $[10 \bar10]$ and $[0001]$ in $x$, $y$ and $z$ directions  of the simulation box, respectively.
The simulation box is kept periodic in all three directions to keep the de-clustered atoms in the box. The minimum distance between the sphere surface and the ends of the simulation box is kept to be at least $25$~\ans\; to avoid the influence of periodic images. The energy is minimised using the conjugate gradient and  FIRE~\cite{guenole:2020} algorithms to maximum force-norm value of $\approx 10^{-5}$~eV/\ans\;. 
% \todo{IMPORTANT!: Info on the orientation and relaxation (to what force norm) of the sample needed} 
%
The particle is initialised with a Gaussian velocity distribution corresponding to $60\%$  of the melting temperature, followed by a microcanonical ensemble for $30$~ps to stabilise the temperature of the system. Once the system is equilibrated at $0.3 \rm T_m$, it is subsequently heated up to the targeted sintering temperature in an canonical (NVT) ensemble for $10$~ps. The NP is then thermalised at the target temperature for $10$~ps. Subsequently, a replica of the particle is created, generating another particle at the same temperature and size, keeping a minimum gap of $50$~\ans{} between them. The entire system is thermalised again at the targeted temperature by an NVT for $10$~ps. The particles are subsequently brought close to each other so that the distance between them is less than the cut-off distance of those interatomic potentials that have such a cut-off. The temperature of the system is then kept at the targeted temperature in an NVT ensemble for $1$~ns. Under favourable conditions, inter-particle diffusion takes place resulting in sintering of the two NPs at constant temperature.

\subsection{Characterisation of sintering}
To compare the performance of the four potentials in terms of sintering it is required to characterise and to quantify the time and temperature dependent sintering process. As a comparison of sintered particles is complex (e.g., in terms of geometry), we use altogether six different  quantities to characterise and quantify sintering, with their evolution providing an indication on the amount of sintering.

 \subsubsection{Shrinkage ratio}
 The coalescence of two particles can be quantified by the shrinkage ratio {\Sf} which is defined as:
\begin{equation}
 S_{\rm f} = 1 - \frac{d_{\rm COM}}{(d_1+d_2)/2} ,	
 \end{equation}
 where $d_1$ and $d_2$ are particle sizes (diameters), and $d_{\rm COM}$ denotes the distance between the centres of mass of them. The latter is obtained as geometric mean of the atomic positions belonging to the particle under consideration. 
{\Sf} is initially negative when NPs are detached, and zero when they are brought into contact. If sintering takes place, the  particles move closer to each other resulting in an increase in {\Sf}.  When the two spheres are completely sintered, the centre-to-centre distance is ideally zero, resulting in {\Sf} being one.

\subsubsection{Surface area}
We note that during sintering, a reduction in surface area is to be expected: this reduction is driven by the movement of surface atoms which are more mobile than the atoms closer to the core of the particle. The transport of material results in reduction of surface area and thus a reduction of overall energy~\cite{ jernot:1981,german:2016}. To compute the surface area, we use the \emph{construct surface-mesh modifier} in \ovito{} ~\cite{stukowski:2009}.  The algorithm constructs a surface mesh with Delaunay tetrahedralisation of the atomic points in conjunction with an appropriate probe sphere radius, where elements are removed that do not fit in the probe sphere. The remaining triangular faces of the tetrahedra defines a 2D boundary or surface of the defined domain. For more details, the reader is referred to~\cite{stukowski:2014}. The virtual probe spheres fill in the empty space between atomic points without spanning the atomic points.  A smaller value gives more details of mesh, and a larger one smooths out the mesh. 
Three values for the probe radius ($2$, $4$ and $6$~\ans\;) were tested. No difference was observed in the results for probe radii of $4$ and $6$~\ans\,; hence a radius of $6$~\ans\; was chosen for the final analysis. The resulting surface area is then normalised by the initial surface area to facilitate comparison of different particle sizes and potentials.

 \subsubsection{Mean square displacement (MSD)}
The MSD defines the average displacement of all atoms in a system with respect to their original positions. It provides a good approximation of diffusion taking place during sintering at different temperatures and is defined as
 \begin{equation} \label{eq:msd_all}  
 {\rm MSD} (t) = \langle \mid \fr_i(t)-\fr_i(0) \mid^2 \rangle,
 \end{equation} 
 where $\fr_i(t)$ and $\fr_i(0)$ are the current and initial positions of atom $i$, respectively.

\subsubsection{Neck curvature}
\begin{figure}
 	\centering
 	\includegraphics[height=13em]{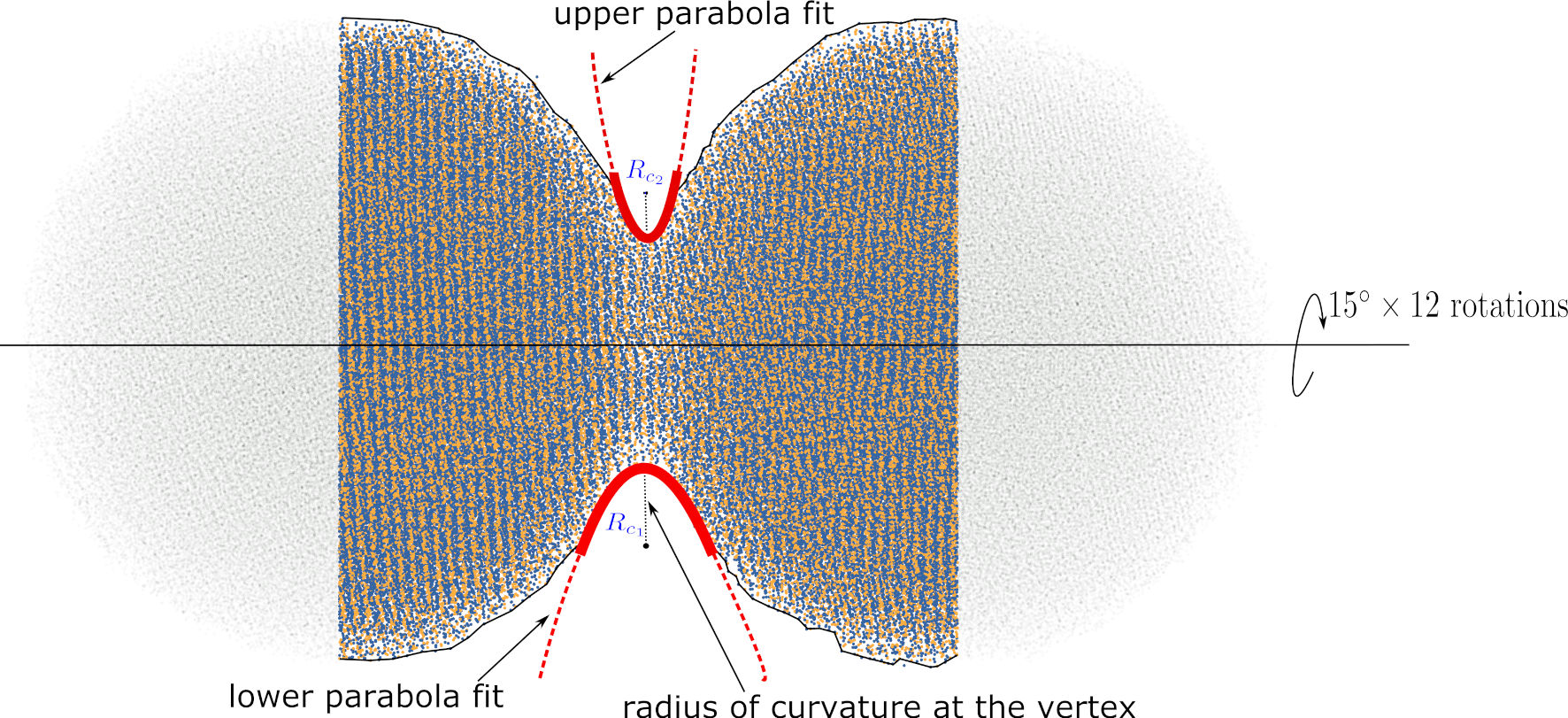}        
 	\caption{%
 		Calculation of neck curvature: Atoms projected onto a plane are shown as coloured points. From them the projected particle boundaries are identified (solid black line). Fitting parabolas to the regions with smallest diameter (red dashed line) allows to compute the radii of curvature, $R_{c_1}$ and $R_{c_2}$. This is repeated for 12 different rotation angles.
 	}
 	\label{fig:schematic_Rc}
 \end{figure}

The transport of surface atoms to reduce the overall surface area results in a reduction of curvature at the neck, i.e., the region where the two particles meet. The neck curvature hence provides a good estimate of the amount of sintering. In the case of completely sintered particles, the neck is completely eliminated, resulting in a value of 0 for the curvature.
We compute the curvature as follows: 3D atomic coordinates are projected onto a plane whose normal is perpendicular to the centre-to-centre axis of the two particles. The boundaries of this region are obtained from spline fitting of the projected data, resulting in two curves that define the profile of the neck. Subsequently, the local minimum/maximum of the two curves are identified, to which second order parabolas ($y = ax^2 + bx + c$ ) %(Eq.~\ref{eqn:parabola}) 
are fitted and from which the curvature can be computed. This is repeated for rotations of $15^\circ$ of the sample about the centre-to-centre axis of the two NPs, resulting in a total of 24 radii of curvatures for a given simulation snapshot. The mean radius of curvature $\bar{R}_{\rm c}$ is  calculated as the arithmetic average of all values from which  the mean neck curvature $\kappa$ is obtained as $\kappa=1/\bar{R}_{\rm c}$. Fig.~\ref{fig:schematic_Rc} provides a schematic diagram of the procedure.

 \subsubsection{Fraction of ions at the neck}
 During the sintering process, the Al and O ions change their positions in the contact region, and this material transport reduces the curvature. 
The change in the geometry of the neck can then also be quantified by the fraction of ions forming the neck region. This is done via the following procedure: The atoms in the initial configuration are assigned an index (0 or 1) based on the NP they belong to. For each snapshot in time, the average index of the nearest neighbour atoms is computed, and those atoms with an averaged index between 0.9 and 1.1 are defined to constitute the neck region.

 \subsubsection{Norm of stretch tensor}
A disadvantage of MSD is that it may lead to erroneous results under the presence of rigid body motion, as will be evident by the results presented below. An alternative measure can be constructed by using displacements whilst accounting for the immediate neighbourhood of individual atoms: we first construct the deformation gradient tensor using the displacement vectors of individual atoms. Using the polar decomposition $\fF_i = \fU_i \fR_i$, we can decompose $\fF_i$ into the right stretch tensor, $\fU_i$, and the rotation tensor, $\fR_i$. Exploiting the orthogonality of $\fR_i$, we obtain $\fU_i$ as:
 \begin{equation}\label{eqn:deform_polar_decom}
 \fU_i = \sqrt {\fF^{\rm T}_i \fF_i},
 \end{equation} 
and subsequently, the rotation tensor as 
 \begin{equation}\label{eqn:deform_polar_decom2}
	\fR_i =  \fF_i \fU_1^{-1}
\end{equation} 
The stretch tensor of individual atoms is subsequently averaged over the nearest neighbors as follows:
\begin{equation}\label{eqn:stretch-tensor_all}
\bar{\fU} = \frac{ \sum\fU_i V_i}{\sum V_i}, 
\end{equation}
where $V_i$ is the atomic volume obtained via a Voronoi construction. For further analysis, we consider the square Euclidean norm of the averaged stretch tensor $\left( \left| \left| \bar{\fU_i} \right| \right|^2 \right)$ as the characteristic that quantifies sintering.

 %%%%%%%%%%%%%%%%%%%%%%%%%%%%%%%%%%%%%%%%%%%%%%%%%%%%%%%%%%%%%%%%%%%%%%%%%%%%%%%%%%%%%%%%%%%%%%%%%%%%%%%%%%%%%%%
\section{Results}
\subsection{Potential and material properties}

\begin{figure}[h]
    \centering
    \includegraphics[width=0.75\textwidth]{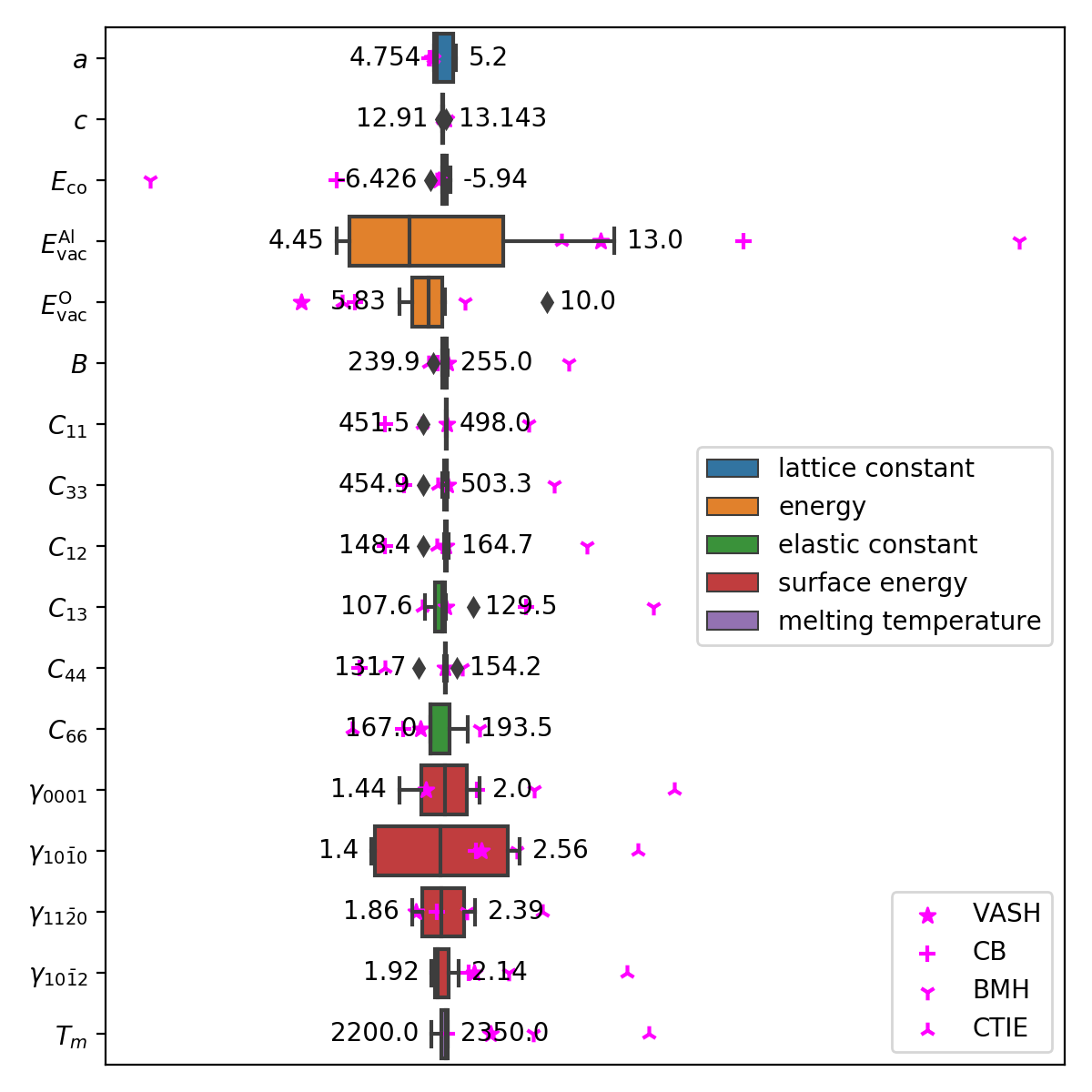}
    \caption{%
        Boxplot of the properties of Al$_2$O$_3$ obtained from experiments together with our data obtained for the four potentials (pink markers). The coloured boxes show the quartiles and median of the data, the ``whiskers'' show the data extent, black diamonds are outliers.
        For plotting all data in the same plot, the data for each boxplot was normalised by the respective mean value; our data was normalised by the mean of the experimental and DFT values. For ease of readability, the values for $C_{14}$ are not shown (cf. Tab.~\ref{tbl:prop_src_simu_all} instead). 
        %
        %\todo{1. change ``K'' to ``B''. 2; Eco, EOvac, all gamma-values differs from table.}  
    }
    \label{fig:exp_MD_DFT_alumina}
\end{figure}

Material properties such as lattice constants, cohesive energies, vacancy formation energies, elastic constants, surface energies and melting temperatures were calculated as described in Section~\ref{sec:MD_method} and the appendix, respectively.
These properties, calculated with all four potentials along with the values from various experiments and density functional theory calculations are provided in Tab.~\ref{tbl:prop_src_simu_all}. A graphical comparison of the range of properties is shown in Fig.~\ref{fig:exp_MD_DFT_alumina}. In the following, we treat values from experiments and DFT equivalently and refer to them collectively as experiments.

\begin{landscape}
    \begin{table}
        \small
        \begin{tabular}{c||c|c||c|c||c|c||c||c|c}
            \hline\hline
            \bf Parameters	    & \multicolumn{2}{|c||}{\bf Experiment \& DFT}	& \multicolumn{2}{|c||}{\bf Vashishta} & \multicolumn{2}{|c||}{\bf Coulomb-Buckingham} & \bf BMH    & \multicolumn{2}{|c}{\bf CTI+EAM}\\  
            {}                  & value    & references                    &  from                  &  our        &  from              &  our                     &  our       &  from                &  our\\  
            {}                  & range    &                    			&  \cite{vashishta:2008} &  results    & \cite{matsui:1996} &  results                 &  results   &  \cite{streitz:1994} &  results\\ \hline \hline 			
            $a$ (\ans))         &$4.754\ldots 5.2$ &{\tiny \cite{lide:2004, gv:1982, ishizawa:1980, hine:2009, lee:1985, shang:2007, liu:2016, tougerti:2011}}         &	            		&$4.72$	        		&$4.71$, $4.7727${\tiny  \cite{sun:2006}}	 	    &$4.68$	    &$4.74$  		&$5.13$					& $4.74$                   	\\  
            %&                       &                    &                   &                     &                  &                           &                       &    \\
            $c$ (\ans))	        & $12.91\ldots  13.143$ &{\tiny  \cite{lewis:1982,  liu:2016, sigumonrong:2011, ishizawa:1980, lee:1985}}              &			            &$13.17$	        		&$13.14$, $12.99${\tiny  \cite{sun:2006}}		    &$13.07$	                               &$13.18$  &    &  $12.98$                            	\\   \hline 
            %&                       &                       &                   &                            &                  &                           &                             &         \\
            $E_{co}$ (eV)	    &$ -6.426 \ldots   -5.94 $ &{\tiny \cite{gieske:1968, hinnemann:2007, rohmann:2011, alavi:2003,  janetzko:2004}}                 &$-6.35$	    	&$-6.35$    		&			                &$-8.91$                                  &$-13.56$       &$-6.36$    & $-6.39$                       	\\   
            %	                &          &	    	&   		&			                &           &$ $                       &                              	     &			        &	  \\ %	
            
            $E^{\rm Al}_{\rm vac}$ (eV)		&$4.45 \ldots  13$ &{\tiny  \cite{hine:2009, matsunaga:2003, mohapatra:1978, carrasco:2004}}                          &			    &$12.56$			        &			                &$16.96$	                                   &$25.46$                            &    &$11.36$   	\\        
            $E^{\rm O}_{\rm vac}$ (eV)		&$5.83\ldots  10$ &{\tiny  \cite{hine:2009, matsunaga:2003, tanaka:2002, carrasco:2004, xu:1997, lei:2013}} 		        &		    	&$3.07$		    	    & 		    	            &$4.57$     	                          &$7.68$                         & 		&$ 4.23$       	\\       
            \hline                                                                                                                                                                                                 %
            $B$ (GPa)  		    &$239.9 \ldots  255$ &{\tiny  \cite{lide:2004, gv:1982, gieske:1968, liu:2016, hine:2009}}                     &$253$		    	&$247$  			&$258$, $391.1${\tiny  \cite{sun:2006}}			                &$255$	        	                    &$377$       &  &$236$                        \\       
            %                    &$253\tnote{7}$                       &                   &                   &                              &                &                           &                                           \\
            $C_{11}$ (GPa)      &$451.5 \ldots  498$ &{\tiny  \cite{lide:2004, gv:1982, shang:2007, gladden:2004, gieske:1968,  hine:2009, goto:1989}}           &$523$		    	&$476$	    		&$714.5${\tiny  \cite{sun:2006}}		            &$458$	    	                           &$659$      &$537$   &$451$                        		\\      
            %                    &$498\tnote{7}$           &              &                   &                   &                   &                   &                   \\
            $C_{33}$ (GPa)  	&$454.9 \ldots  503.3$ &{\tiny  \cite{lide:2004, gv:1982, gladden:2004, gieske:1968, hine:2009, goto:1989, shang:2007}}          &$427$			&$415$		    	&$709.3${\tiny  \cite{sun:2006}}		            &$490$  		                         &$712$       &$509$ &$483$                           	\\        %      &$509$		   	    &$ $			&$486$			&$ $               &$ $  		&$ $
            $C_{12}$ (GPa)  	&$ 148.4 \ldots  164.7$ &{\tiny  \cite{lide:2004, gv:1982, gladden:2004, gieske:1968, hine:2009, goto:1989, shang:2007}}	     &$147$			&$155$		    	&$327.5${\tiny  \cite{sun:2006}}		            &$172$	    	                  &$254$         &$180$ &$157$                     		\\        %      &$180$	       		&$ $			&$134$			&$ $               &$ $		&$ $

            $C_{13}$ (GPa)  	&$107.6\ldots  129.5$ &{\tiny  \cite{lide:2004, gv:1982, gladden:2004, gieske:1968, hine:2009, goto:1989, shang:2007}}	         &$129$			&$135$		    	&$187.8${\tiny  \cite{sun:2006}}		            &$124$	    	                 &$213$              &$106$  &$106$               		\\        %      &$106$	       		&$ $			&$145$			&$ $               &$ $		&$ $
            
            $C_{44}$ (GPa)  	&$131.7\ldots  154.2$ &{\tiny  \cite{lide:2004, gv:1982,  gieske:1968, hine:2009, goto:1989, shang:2007}}             &$135$			&$149$		    	&$99.7${\tiny  \cite{sun:2006}} 		            &$117$	    	                          &$157$             &$130$  &$112$           		\\        %      &$130$	       		&**		    	&$164$ 			&$ $               &$ $		&$ $

            $C_{66}$ (GPa)  	&$167\ldots  193.5$ &{\tiny  \cite{lide:2004, gv:1982, gladden:2004, gieske:1968, hine:2009}}         &$174$			&$160$		    	&$193.5${\tiny  \cite{sun:2006}} 		            &$148$	    	                  &$202$       &$179$  &$147$                      		\\

            $C_{14}$ (GPa)  	&$-24 \ldots  +22.5$ &{\tiny \cite{gladden:2004, gieske:1968, hine:2009, goto:1989, shang:2007}}           &$7.5$							&$9$	            &$-42.2${\tiny  \cite{sun:2006}}		            &$0.02$	    	                          &$0$               &$-30$  &$-0.001$               		\\       \hline
            $\gamma_{0001}$ (J/m$^2$)	   &$1.44 \ldots  2.00 $ &{\tiny  \cite{zhou:2004, manassidis:1994, marmier:2004, sun:2006dft, pinto:2004, tougerti:2011, kurita:2010}}	         &	    &$1.63^{Al}$		       		&		        	    &$1.98^{Al}$	                       &$2.38^{Al}$        &$2.67$  &$3.36^{Al}$                     		\\    
            
            %		$\gamma_{1\bar100}$ ($J/m^2$)	&$2.65\tnote{32}$           &	&$2.26^{Al-O}$		       		&		        	    &$2.17^{Al-O}$	                          &$2.48^{Al-O}$                           &  &$3.48^{Al-O}$        	\\    
            
            %		$\gamma_{1\bar210}$ ($J/m^2$)	&           &	&$1.89^O$		       		&		        	    &$2.07^O$	                               &$2.32^O$	                  &  &$2.96^{O}$             	\\    
            
            $\gamma_{10\bar10}$ (J/m$^2$)	&$1.40 \ldots  2.56$ &{\tiny  \cite{zhou:2004, manassidis:1994, marmier:2004, sun:2006dft}}       &	&$2.26^{Al-O}$		       		&$ $		    	    &$2.22^{Al-O}$	       	                       & $2.54^{Al-O}$	                   &$1.28$  &$3.49^{Al-O}$            	\\   
            
            $\gamma_{11\bar20}$ (J/m$^2$)	&$ 1.86 \ldots  2.39 $ &{\tiny  \cite{ zhou:2004, manassidis:1994, marmier:2004, sun:2006dft, tougerti:2011, kurita:2010}}     &	&$1.89^O$		       		&$ $		    	    &$2.07^O$	                       &$2.32^O$        	              &$1.81$  &$2.96^{O}$         	\\   
            
            $\gamma_{10\bar12}$ (J/m$^2$)	&$1.92 \ldots 2.14$ &{\tiny  \cite {zhou:2004, manassidis:1994, sun:2006dft}}   		  &			&$2.26^{Al-O}$		       		&$ $		    	    &$2.22^{Al-O}$	                               &$2.54^{Al-O}$	                       &$1.80$  &$3.49^{Al-O}$        	\\    \hline
            
            $T_m$ (K)             &$2200 \ldots  2350 $ &{\tiny  \cite {shen:1995, shen1:1995, sakate:1995}} 	      &$2760$            &$2740$                 &$2425-2475${\tiny  \cite{ahuja:1998}}          	            &$2340$                                 &$3140$              &  &$4200$                \\
            \hline\hline
        \end{tabular}
        
        \caption{\footnotesize%
            Summary of our obtained material properties in comparison to values reported in the literature. The superscripts $Al$, $Al-O$ and $O$ of the surface energies denote the terminating atoms on the slab surfaces. A simple check ($C_{66}=(C_{11}-C_{12}) /2$) for each potential provides an additional validation of elastic constants calculation. 
           % To validate the data for elastic constants, calculated from simulations, a simple check ($C_{66}=(C_{11}-C_{12}) /2$) for each potential can be performed. 
            %
            %\todo{Why are there significant differences between our values for CTIE and the original publication?}. 
            %\Stefan{is it a good place for the following comment here in the caption?}
            %\Arun{Yes, I think so. But the value of 2.3 J/m2 in the caption differs from the value of 1.28 in the table. If we are unsure, it would be probably better to leave it out. Seems inconsequential for the results}
            %
            %%% 
			%            Note, that the slab-thickness for calculating surface energies in~\cite{streitz:1994} was not clearly mentioned. However, \citep{streitz:1994} provided the changes in surface energies for the $(10\bar10)$ surface which converged to value of $2.3 (J/m^2)$. The values in the table for surface energies do not seem to be converged values.
		}
        \label{tbl:prop_src_simu_all}
    \end{table}
\end{landscape}
For most properties, the values predicted by almost all interatomic potentials are well within or close to the range of values obtained from experiments (cf. Tab.~\ref{tbl:prop_src_simu_all}). The notable exception is seemingly the \potB{} potential which predicts values for 0~K properties that are significantly higher than those obtained from experiments. Exceptions are also seen in the case of vacancy formation energy of Al; while \potV{} and \potCT{} predict values towards the higher values observed in experiments, values from \potC{} and \potB{} are significantly beyond the range of experimental values. The latter two also predict significantly higher (numerically, lower) cohesive energy values than that observed from experiments. In case of the surface energies, we observe values that are by and large close to the range of data obtained from experiments. Only the \potCT{} is an exception here: energies for all the surfaces are significantly higher than that seen in experiments.

The predicted melting temperature values obtained via MD simulations with the four potentials show that only the \potC{} potential predicts a value well within the range of experimental data. The values from all other potentials are significantly higher, with the highest value of 4200~K by the \potCT{} potential. To facilitate objective comparison of the sintering process with the different potentials, we use the homologous temperature of the corresponding potential.

In Fig.~\ref{fig:exp_MD_DFT_alumina} we show the 25/75 percentile, median and total range of material parameters measured or calculated by various techniques. Outliers are detected based on 150\% of the inter-quartile range. The figure shows that some of the obtained material properties exhibit significant scatter together with a strong skewness of the distribution. For example, the highest melting temperature ($4200$~K) is almost $210\%$ more than the lowest melting temperature ($2000$~K) reported in experiments; the median of $E_{\rm vac}^{\rm O}$ is strongly shifted towards the left, and all of our computed values for this property are outside the 25/75 percentile range of the literature values. The literature values for the lattice constants, cohesive energy as well as most of the elastic properties don't show much variance. Looking at this figure, the best conformance with the experimental values for most of the properties %, but for the vacancy formation energy of Al,% Stefan: I still don't understand this sentence part...
is obtained with the \potC{} potential. %\todo{last sentence is not clear}

\subsection{Sintering} 
\begin{figure}%[htbp!] %[htbp!]
	\centering
	\includegraphics[width=0.92\textwidth]{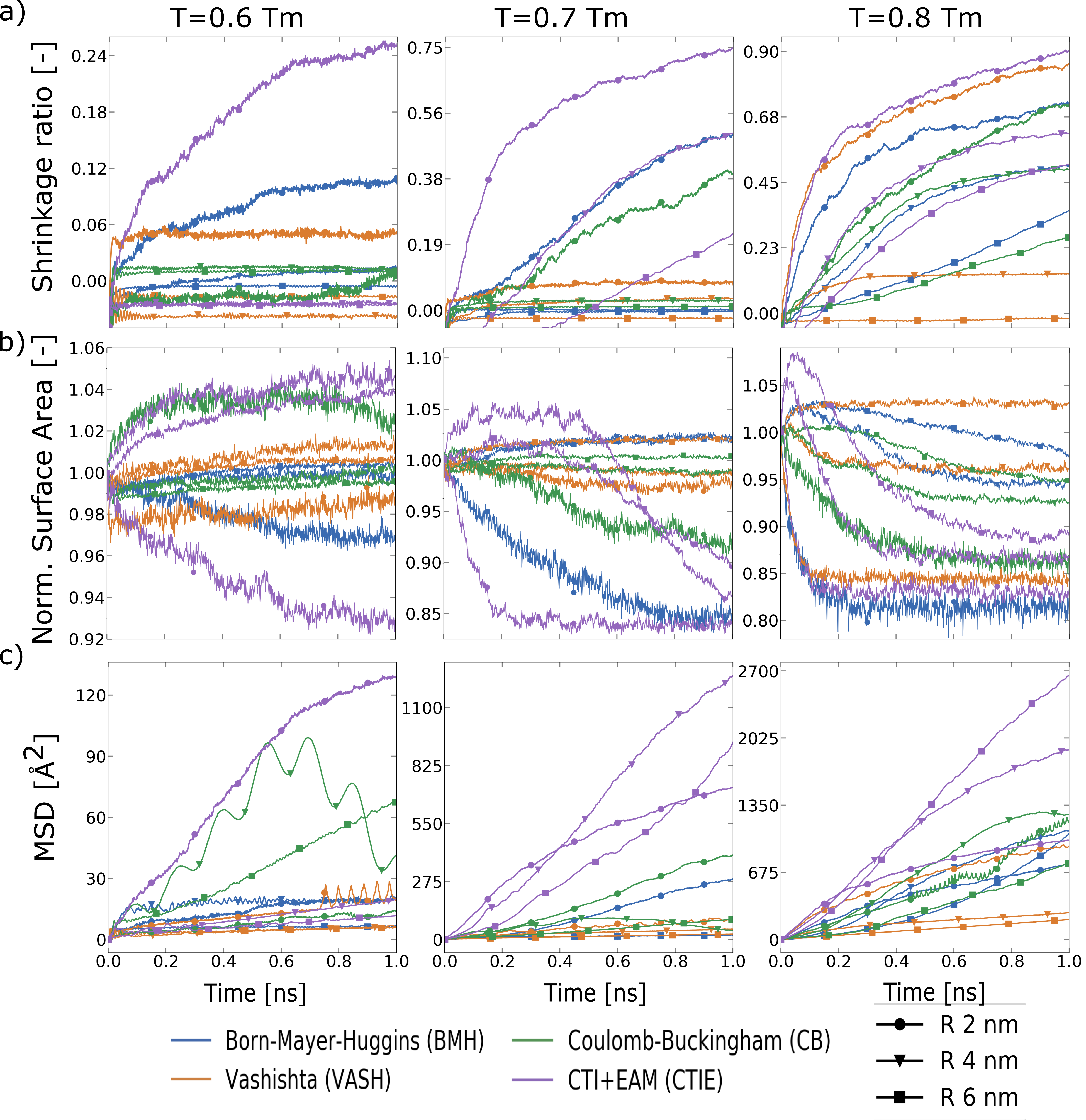} 
	\caption{Evolution of the first three global quantities: a) shrinkage ratio, b) normalised surface area, and c) MSD. The left, middle and right column are for  $60\%$, $70\%$ and $80\%$ of melting temperatures. Markers are shown at every $20$ data points (= every $2$~ps).}% \todo{Shyamal, please double check the text and make sure that it refers to the correct Figs/subplots.}}
	\label{fig:global_quantities_246nm1}
\end{figure}

\begin{figure}%[htbp!] %\ContinuedFloat
	\includegraphics[width=0.92\textwidth]{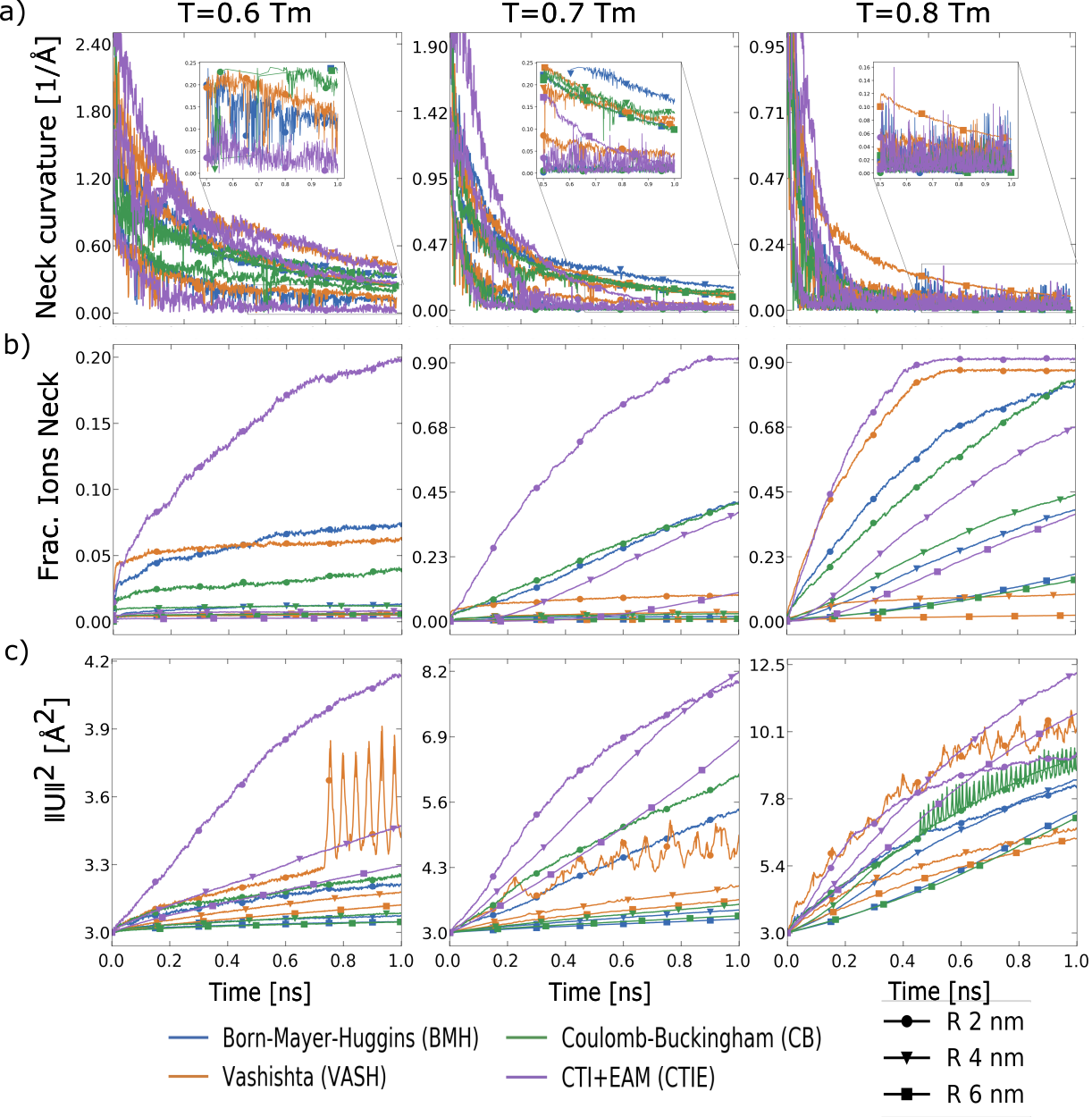}
	\caption{Evolution of the last three global quantities:   a) curvature at the neck, b) fraction of ions at the neck, c) norm of the stretch tensor. The left, middle and right column are for  $60\%$, $70\%$ and $80\%$ of melting temperatures. Markers are shown at every $20$ data points (= every $2$~ps). }
	\label{fig:global_quantities_246nm2}
\end{figure}

%\todo{The right reference to the movie numbers goes in here!} 
The results of the sintering simulations for each investigated temperature and particle size are shown as supplementary movies M1 to M3, with each movie comparing the prediction of all four potentials. 
Fig.~\ref{fig:global_quantities_246nm1} and~\ref{fig:global_quantities_246nm2} show the evolution of all investigated characteristics over time for all four potentials (indicated by the different line colours), for three different particle sizes (indicated by markers) and for different temperatures (sub-plots in columns).

\paragraph{Shrinkage ratio:}
Fig.~\ref{fig:global_quantities_246nm1}a shows the evolution of the shrinkage ratio ({\Sf}). For all four potentials, we generally observe an increase of {\Sf} with time, particularly at higher temperatures, implying that the two particles move closer to each other. However, the magnitude of this movement differs strongly from one potential to another, for all particle sizes and temperatures. At {\Ta}, for {\Rb} and $6$~nm, {\Sf} remains negative or close to zero throughout the simulation, indicating that the two particles barely come into contact with each other. At this temperature, a non-negligible value of \Sf{} is only observed for particles with \Ra{} and potentials \potCT{}, \potB{} and \potV; while the former two display a steady increase in \Sf{}, \potV{} shows an almost instantaneous increase to $\approx5$\% at the start which is then maintained throughout the simulation.

The evolution of \Sf{} at \Tb{} follows similar trends, however, for NPs with \Ra{}, a non-negligible positive \Sf{} is seen for all four potentials. While a steady increase over the entire simulation time and a value greater than 40\% is observed for \potB, \potC{} and \potCT{}, \potV{} displays an increase only within the first 0.2~ns and a saturation to significantly lower value of roughly $10\%$. Additionally, \potCT{} shows a \Sf{} value of roughly 50\% and 20\% for \Rb{} and 6~nm, respectively; for these particle radii, other potentials do not show any noticeable value of \Sf{}.

Significantly different trends are seen at the higher temperature of {\Tc}. Both {\potC} and {\potB} show very similar characteristics in the evolution of {\Sf} over time. For the three particle sizes, {\Sf} evolves towards higher values than at lower temperatures and attains values of $0.7$, $0.4$ and $0.2$ after $1$~ns, evidencing a trend of decreasing {\Sf} with increasing particle size.  In general, \Sf{} with \potCT{} is higher than that predicted by other potentials. Very contrasting trends are seen for the {\potV} potential: Whist the evolution for \Ra{} follows that of \potCT{}, for \Rb{}, \Sf{} saturates to a value of 0.1 after 0.2~ns and remains negative for \Rc{}.

\paragraph{Surface area:} 
The evolution of overall surface area (\Af{}) of the NPs is shown in Fig.~\ref{fig:global_quantities_246nm1}b. To enable easy comparison, the values are normalised w.r.t. the corresponding initial area before the start of the sintering process. Therefore, values should start at 1 and are expected to reduce during the simulation time if sintering proceeds favourably.

By and large, evolution of \Af{} displays trends similar to that observed for {\Sf}: when the two NPs come close to each other, a decreasing surface area over time is generally observed (compare at \Ta{}: \potB{}, \potV{} and \potCT{} potentials and at \Ra{} and \Tb{}: \potB{}, \potC{} and \potCT{}). For cases where no sintering it expected (evidenced by zero shrinkage ratio), \Af{} remains close to unity.

%\todo{Arun said: "For \Tb{} and \Tc{}, we see an increase in \Af{}. We had already discussed about this and if I remember correctly, further simulations were performed. This needs to be incorporated.". Also: "to be confirmed: Increase of less than 10\% in the shrinkage ratio might not indicate sintering"}
%\discuss{The max decrease in the surface area in the ideal case would be 79.37 \%}

\paragraph{Mean square displacement: }
The evolution of MSD over time is shown in Fig.~\ref{fig:global_quantities_246nm1}c. The curves clearly show an increase in MSD over time, albeit with different amounts for different sample sizes and potentials. With increasing temperature, a significantly higher MSD is observed, indicating the propensity for increased diffusion at higher temperatures. However, the trends do not complement those observed with \Sf{} and \Af{}. For example, the MSD for \potCT{} at \Tb{} is greater than the MSD for all other potentials and particle sizes, which contradicts the trends observed with \Sf{}. This can be observed at \Ta{} and \Tb{} as well. Furthermore, the evolution of MSD with the \potC{} potential at \Ta{} shows non-monotonic evolution with periodic crests and troughs, indicating artifacts associated with the superposition of rigid body motion.

\paragraph{Neck curvature: }
The change in the neck curvature $\kappa$ with time is shown in Fig.~\ref{fig:global_quantities_246nm2}a and, as expected, decreases with time. A value of zero indicates a complete removal of the neck region. The initial value of $\kappa$ for larger particles are higher than that of the smaller particles, and at the same time, the initial change in the curvature for \Ra{} is faster than that for \Rb{} and \Rc{} for all temperatures. Generally, no unexpected trends are observed and the evolution is consistent with the trends observed for \Sf{}.

\paragraph{Fraction of ions at the neck:}
The change in the accumulated fraction of ions at the neck, N$_f^i$, is shown in  Fig.~\ref{fig:global_quantities_246nm2}b. The observed trends here align well with those seen in the shrinkage ratio. With increasing temperature, N$_f^i$ reaches values as high as 0.9, indicating that almost the entire system of two particles is now classified as the neck region. Such a scenario occurs when the two particles have completely sintered and remixing of atoms from the two particles has taken place. At lower temperatures N$_f^i$ reaches substantially lower values; the only exception seemingly is for \Ra{} with the \potCT{} potential at \Tb{}, where a value of more than 0.9 is reached.

 \paragraph{Norm of stretch tensor: }
Fig.~\ref{fig:global_quantities_246nm2}c shows that the evolution of {\Un} starts at a value of $3$ since the initial stretch tensor is an identity tensor. The square of the norm is used here to facilitate comparison with MSD. Here, it is evident that there are  differences between the evolution of \Un{} and MSD: at \Tc{}, the MSD after 1~ns for \potV{} is significantly lower than other potentials for \Rb{} and 6~nm, whereas the \Un{} values are very similar to those from \potB{} and \potC{} for \Rc{}. Furthermore, the non-monotonic evolution with periodic crests and troughs seen with the \potC{} potential at \Ta{} vanishes in the evolution of \Un{}. However, for the \potV{} potential, the evolution of \Un{} shows fluctuations from the mean at all temperatures. These fluctuations, also visible with the evolution of MSD.

 %%%%% Analysis and discussion of the gloabl qountaties %%%%%%%%%

\section{Discussion}\label{sec:discussion} 	

 We study the different formulations and parameterisation of four potentials and investigate how they result in different amounts of sintering for three different temperatures and particle sizes. In the current work, six measures are used to characterise sintering and to quantify the process. It is hence important to discuss the pros and cons of these measures, since not all measures are able to quantify the sintering process to the same extent.
 
\begin{figure}%[ht!] %[htbp!]
    \centering
    \includegraphics[width=0.99\textwidth]{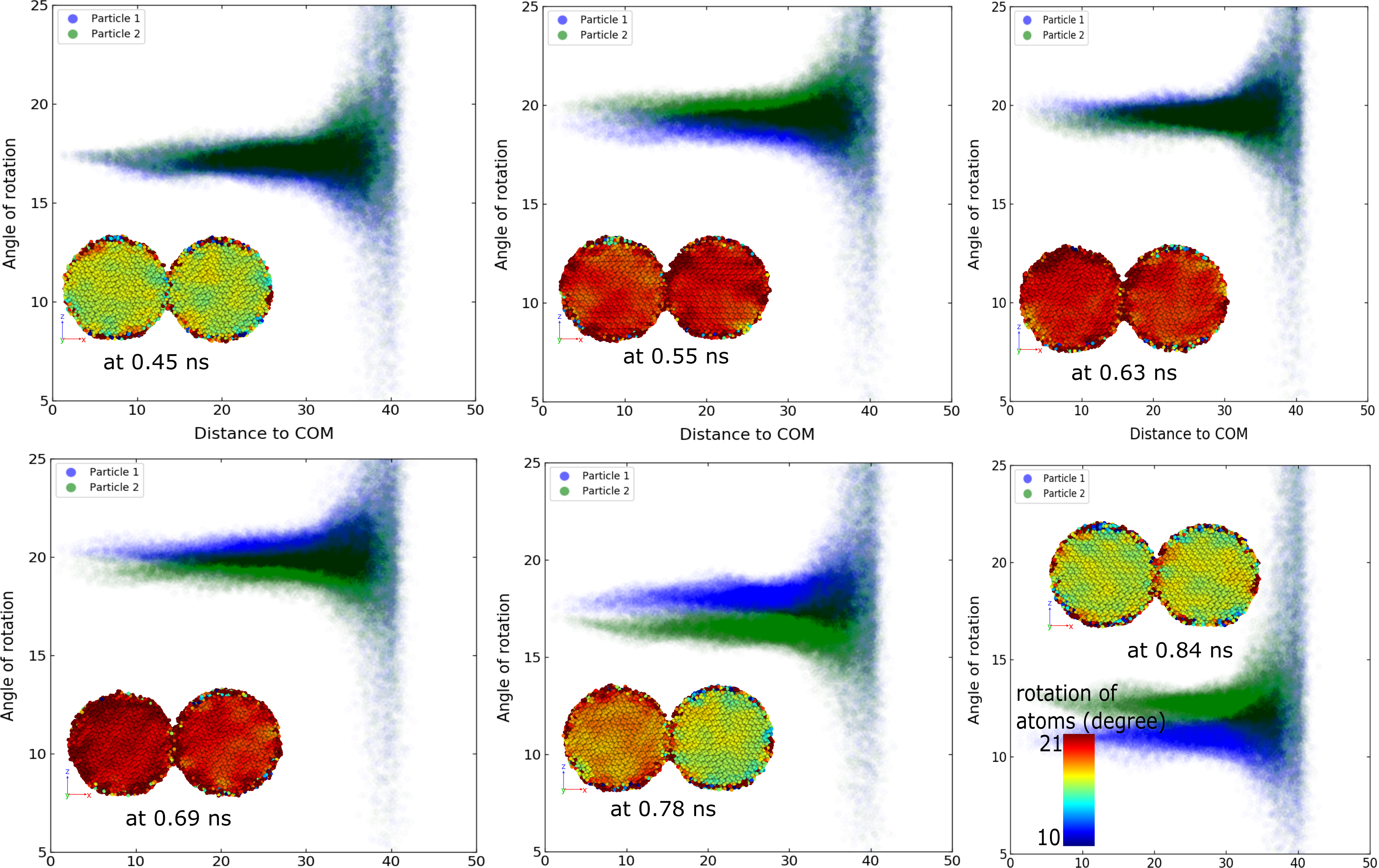} 
    \caption{Rotation distribution of atoms for two NPs with the distance from centre of masses of the corresponding particles is plotted. The insets are snapshots of atomic structure contoured with the amount of rotation at various simulation times with respect to their positions at time zero for potential {\potC} and radius {\Rb} at {\Ta}. The distribution and the snapshots are chosen at the simulations times when MSD changes its evolution path (see Fig.~\ref{fig:global_quantities_246nm1}c).}
    \label{fig:atom_rotate}
\end{figure}

In particular, it is visible that the MSD is prone to influences from rigid body motion, resulting in a non-monotonic evolution over time as seen for instance with \potC{} for \Rb{} and \Ta{}. We note that the simulations are performed with periodic boundary conditions and the MSD is calculated using only the atoms participating in the sintering process, i.e. atoms that detach themselves from the main particle are neglected. To understand the non-monotonic evolution in MSD, we investigate the rigid body rotation of the two particles, see Fig.~\ref{fig:atom_rotate}. 
This is done by using the polar decomposition of the deformation gradient $\fF=\fR\fU$ and expressing the rotation tensor $\fR$ in terms of an angle that denotes the rotation of an individual atom from the initial configuration. It is evident that the crests and troughs in the evolution of MSD correspond to significant changes in the rotation of the two particles. This non-monotonic evolution is eliminated when the rotational part is removed from the displacement, as is the case with the stretch tensor \Un{} (cf. Fig.~\ref{fig:global_quantities_246nm2}c). MSD, as a quantifying measure of the sintering process, must hence be used with caution. Indeed, we strongly suggest the use of \Un{}, which essentially encapsulates MSD whilst accounting for the immediate neighbourhood of individual atoms.
 
All six measures aim to quantify the same sintering process. Therefore, they should be able answer the questions a) which of these quantities give a definitive indication on the amount of sintering?, and b) what numerical value of the said quantity indicates complete sintering? In principle all six measures should be consistent with each other if they perform equally well. However, significant discrepancies can be observed in the evolution of these quantities. For instance, MSD (and also \Un{}) for \Rb{} and \Rc{} at \Tc{} for \potCT{} is higher than that for all other potentials and particle sizes. This trend is not observed with other quantities, such as the fraction of ions in the neck, \ionF{}, or the shrinkage ratio, \Sf{}. Also, it is not completely evident from the quantities themselves if we indeed have complete sintering or not. 

\begin{figure}%[ht!] %[htbp!]
	\centering
	\hbox{}\quad
	\begin{minipage}[c]{0.39\textwidth}
	    \small
	    \textbf{Interpretation of the colour code:}
	    \footnotesize
	    \begin{itemize}\setlength{\itemsep}{0pt}%
	    \item [0:] no substantial change relative to the initial configuration
	    \item [1:]  minimal formation of the neck region
	    \item [2:] significant progress of the neck region
	    \item[2.5:] sintering of the two particles without remixing of ions
	    \item[3:] sintering with complete remixing.
	    \end{itemize}
	\end{minipage}
	\hfill
	\begin{minipage}{0.5\textwidth}
	    \includegraphics[width=\textwidth]{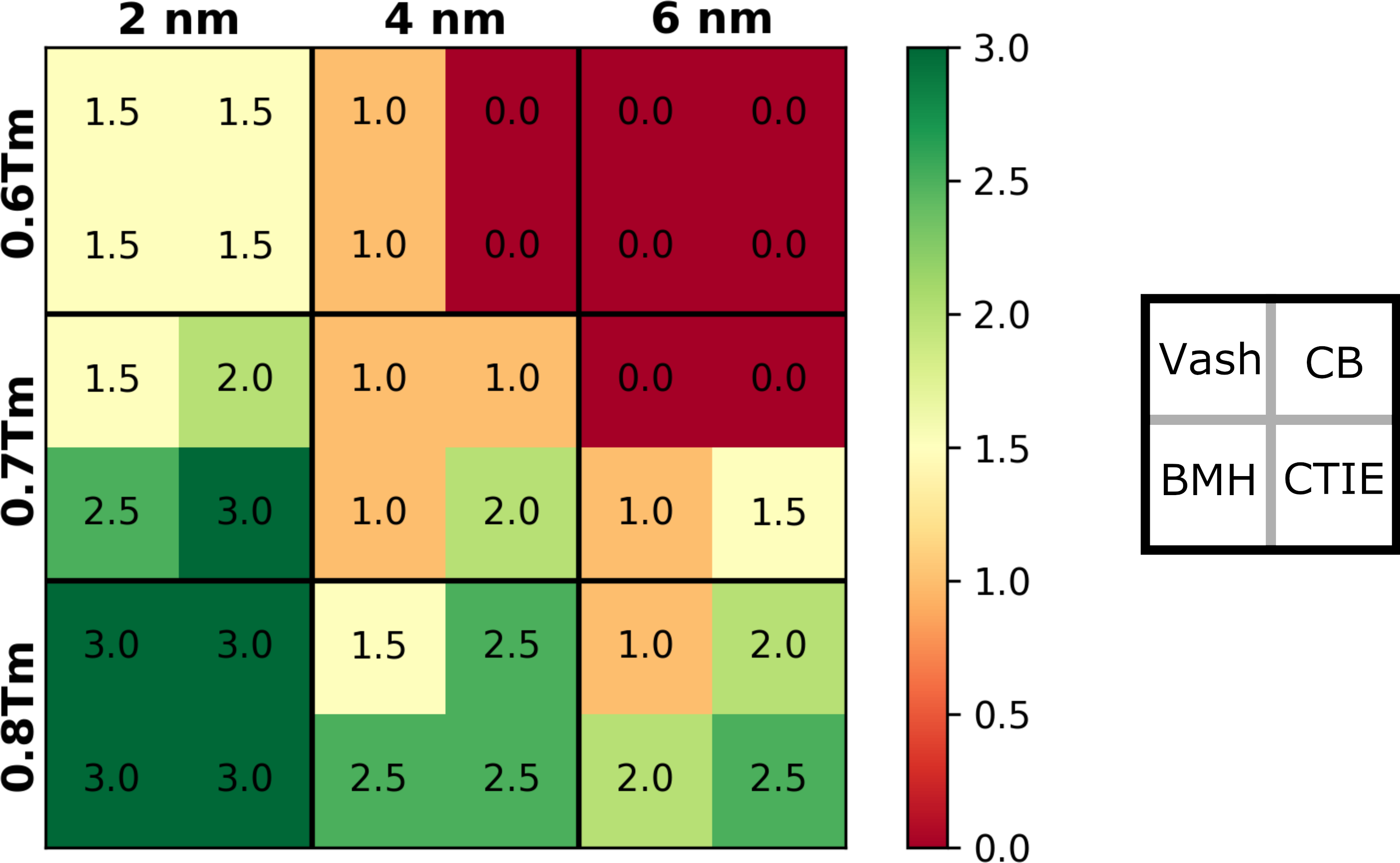}
	\end{minipage}
	%\hfill\hbox{}
	\caption{%
        Quantification of the amount of sintering by ``visual inspection'' for all investigated potentials,  particle sizes and  temperatures.
    }
	\label{fig:SinteringProgress}
\end{figure}

How strongly do the six sintering measures correlate with our expectation and experience? Fig.~\ref{fig:SinteringProgress} provides a quantification of the amount of sintering obtained via visual inspection. The scale of \mbox{0--3} indicates no substantial change in the initial configuration (0) to complete sintering (3).

The normalised surface area \Af{} and the neck curvature are perhaps the most intuitive measures to set numerical thresholds that indicate complete sintering. In the ideal case of two particles being completely sintered and become one sphere, the normalised surface area should decrease to a value of approx. 0.79, and the neck curvature to zero. However, the discrete nature of the system and the approximations used in the two algorithms can result in large fluctuations due to small perturbations, making these quantities sensitive and error-prone. 

The shrinkage ratio \Sf{} and the fraction of ions in the neck \ionF{} are less intuitive, but seemingly the more robust measures to quantify sintering. In the ideal case of complete sintering, \Sf{} should evolve to a value of 1. In the way how we define the neck region, complete sintering would involve remixing of atoms from the two original NPs resulting in the entire sintered particle being classified as neck leading to a value of 1 for \ionF{}. In practice, however, values of approximately 0.7 and 0.8 for \Sf{} and \ionF{}, respectively, are seemingly sufficient to indicate complete sintering, while values of 0.5 and 0.35, respectively, indicate partial sintering, i.e., two particles have become a single particle, albeit without significant remixing of ions from the two particles. The latter case is merely a question of kinetics and would result in the former case with increased simulation time. We note, nonetheless, that these numerical values have no physical meaning and are mere observations from our simulations.

A focal point of the current work is the comparison of interatomic potentials for the simulation of sintering. A key component of atomistic simulations is the reliability and applicability of interatomic potentials. Empirical potentials, such as those used in the current work, are usually derived by fitting a functional to certain properties of the material under consideration. All four potentials used in the current work make use of lattice constants and the bulk modulus of alumina in their fitting procedure. Cohesive energy is used additionally for the parameterisation of \potCT{} and \potV{} potentials. In this regard, it is unsurprising that the values of these properties fall well within the range of values obtained from experiments (see Fig.~\ref{fig:exp_MD_DFT_alumina}). All other material properties listed in \tref{tbl:prop_src_simu_all}, including the vacancy formation energies and the surface energies, are predicted values and differ significantly from one potential to another.

The decision on which interatomic potential to use is generally taken based on the accuracy of the predicted material properties, which are usually not a part of the fitting procedure. The two main properties of interest in the sintering process are the surface energies and the vacancy formation energies. The former is of lesser relevance in the current work due to the usage of spherical particles. Nevertheless, we note that all potentials but for \potB{}, predict surface energies within the range determined by experiments.

The values of the vacancy formation energies -- both Al and O-site vacancy energies -- predicted by the four potentials are, however, significantly different from one another. While the values of the O-site vacancy energies are very close to experimental values, those of Al-site vacancy energies predicted by \potC{} and \potB{} potentials are significantly higher than experimental values. In general, the following trend is observed in the predicted values of the vacancy formation energies: \potV{} $<$ \potCT{} $<$ \potC{} $<$ \potB{}, with the lowest values predicted by the \potV{} potential.

Given the proximity in the values of most material properties to those obtained from experiments, and the low vacancy formation energies, we should expect the best prediction of the sintering process with the \potV{} potential. However, comparison of the amount of sintering (see Fig.~\ref{fig:SinteringProgress}) shows completely contradictory results with the least amount of sintering with \potV{} and furthermore, a near absence of sintering of particles of larger size (\Rb{} and \Rc{}) even at \Tb{} and \Tc{}. By contrast, all other potentials show an increased amount of sintering at \Tc{}. The reduced progress in sintering for the highest particle radius of \Rc{} reflects the slower kinetics, and is likely to result in a completely sintered state with increased time. With the \potV{} potential, however, an increased simulation time is unlikely to yield any changes as is evident in the characteristics presented in Fig.~\ref{fig:global_quantities_246nm1} and Fig.~\ref{fig:global_quantities_246nm2}. 

It hence follows that the origin of these differences lies in the functional form and treatment of the individual potentials. The \potV{} potential includes both two-body and three-body interactions, the latter being applied only to triplets of atoms. The former, which includes Coulomb interactions, is truncated at $r_{\rm cut}=6~\AA$ \cite{vashishta:2008}. Such truncation is unlikely to have an effect on periodic systems as is evident in the accuracy of the material properties predicted by the potential. In the presence of free surfaces, such truncation results in significant discrepancies like pseudo size effects observed in the sintering of larger particles at higher temperatures.

As pointed out in \sref{sec:potentials}, the formulations of the \potB{} and \potC{} potentials are effectively identical in the manner they are used in the current work. The \potB{} potential is indeed a re-parameterisation of the \potC{} potential \cite{bouhadja:2013}. The key difference between the two potentials is in the effective charges ($Z_i$) used for Al and O ions: in the \potB{} potential charge values close to those in pure oxide systems \cite{vanBeest:1990} are used. This difference in charges has a significant effect on the material properties predicted by the two potentials, but has apparently no effect on kinetics of the sintering process. Evidently, for all particle sizes and temperatures considered in the current study, the amount of sintering predicted by the two potentials is almost the same.

Of the four potentials, the \potCT{} potential results in the highest kinetics of the sintering process. A plausible reason for this is the charge equilibration performed in this potential. In the presence of a heterogeneous electrostatic environment, as e.g. interfaces and free surfaces, ions are likely to have varying charges depending on their local environment. The ability to perform charge equilibration allows for different charges for different ions, depending on their local environment, which we believe is the reason for improved kinetics with the \potCT{} potential. A further improvement in the simulations can be introduced by activating dynamic charge transfer in the \potCT{} potential since this is known to enhance atomic diffusion in surface regions of nanoparticles \cite{ogata:2000}.

We hence summarize that accurate treatment of Coulomb effects, together with correct effective charges of the ions, are important for effective prediction of sintering in atomistic simulations of alumina. The usage of effective charges can result in improved kinetics of the sintering process. Truncation of Coulomb effects, however, is more critical and can result in spurious size effects during sintering.

\section{Conclusions}
In this study, we perform atomistic simulations of sintering and compare four interatomic potentials for alumina. The four potentials -- \potV{}, \potC{}, \potB{} and \potCT{} -- differ in the formulation as well as in their treatment of Coulomb interactions and individual charges. The potentials are first compared in terms of their ability to predict fundamental material properties. Three different particle sizes and temperatures are used to study the sintering process, which is subsequently characterised using six different quantities. The findings of the study can be summarised as follows:

\begin{itemize}
\item All four potentials predict lattice constants that are well within the range of experimental values. Only \potV{} and \potCT{} predict cohesive energies that are close to experiments. Both \potC{} and \potB{} predict significantly higher values of cohesive energy
\item The values of most elastic constants are significantly higher with the \potB{} potential. With the other three potentials, most of the values are very close to those obtained from experiments.
\item The surface energy values predicted by the \potCT{} potential are significantly higher than the experimental values. All other potentials predict values well within the range determined by experiments.
\item The vacancy formation energy values of both Al and O sites follows the following trend: \potV{} $<$ \potCT{} $<$ \potC{} $<$ \potB{}. 
\item All potentials show an increasing amount of sintering with increasing temperature for the nanoparticle with \Ra{}, with complete sintering observed at \Tc{}.
\item Very little or no sintering is observed for the largest particle size of \Rc{} at temperatures \Tb{} and below.
\item Except for the \potV{} potential, all other potentials predict significant amount of sintering at \Tc{} for all particle sizes.
\item For the quantification of the sintering process, the use of \Un{} is recommended over MSD, since the latter is particularly prone to influences of rigid body motion of the particles.
\item The shrinkage ratio (\Sf{}) and fraction of ions in the neck (\ionF{}) provide the best quantification of the sintering process. Numerical values of 0.7 and 0.8 for \Sf{} and \ionF{}, respectively, indicate complete sintering.
\item The abrupt cutoff of Coulomb interactions is deemed as the reason for the spurious behaviour of the \potV{} potential, which shows almost no sintering for particle sizes above \Ra{}, even at the high temperature of \Tc{}.
\item The \potCT{} potential predicts the highest amount of sintering amongst all potentials compared in the current work. The increased kinetics is attributed to the ability to perform charge equilibration within the potential.
\end{itemize}

%%%%%%%%%%%%%%  ACKNOWLEDGMENT   %%%%%%%%%%%%%%%
\section*{Acknowledgments}\label{SEC:ack}
The authors thank Prof. H. Riedel for fruitful discussions. This work is supported by the German Research Foundation (DFG) within theframework of the Collaborative Research Center SFB 920 Project-ID169148856 ``Multi-Functional  Filters  for  Metal  Melt  Filtration -- A Contribution toward Zero Defect Materials'', subproject B04. The authors gratefully acknowledge (i) computing time granted on the high-performance compute cluster operated by the University Computing Center (URZ) of the TU Bergakademie Freiberg and (ii) computing time granted through JARA on the supercomputer JUQUEEN~\cite{jureca} at Forschungszentrum Jülich.

%%%%%%%%%%%  REGARDING DATA  %%%%%%%%%%%%%
\section*{Data availability}
Postprocessed datasets and chosen visualisation scripts  can be obtained from gitlab XYZ, raw data and simulation/analysis code is made available upon reasonable request.

\bibliography{lit_ok2}

%\newpage
\section*{Appendix}
%%% %%%% THE PARAMETERS OF THE POTENTIALS %%%%%%%%%%%%%%%5
\begin{center}
	\begin{table}
		\begin{tabular}{|c|c|c|c|c|c|c|c|}
			\hline
			Vashishta~\cite{vashishta:2008} 										&Al		&O		&Al-Al		&Al-O		&O-O		&Al-O-Al		&O-Al-O				\\
			\hline  
			$Z$ (e) 											&$1.5237$  	&$-1.0158$		&		&		&		&		&				\\
			$\eta_{ij}$ 								&		&		&$7$		&$9$		&$7$		&		&				\\
			$H_{ij}$ (eV \ans{arg1}) 	&		&		&$12.7506$		&$249.3108$		&					&		&		\\
			$D_{ij}$ (eV \ans{arg1}$^4$) 	&		&		&$0$		&$50.1522$		&$44.5797$					&		&		\\
			$W_{ij}$ (eV \ans{arg1}$^6$) 	&		&		&$0$		&$0$		&$79.2884$			&				&		\\
			$B_{ijk}$ (eV)  						&		&		&		&		&			&$8.1149$		&$12.4844$				\\
			$\bar{\theta}_{ijk}$ (deg)  						&		&		&		&		&			&$109.47$		&$90.0$				\\
			$C_{ijk}$ 						&		&		&		&		&			&$10$		&$10$				\\
			$\gamma$ (\ans{})  						&		&		&		&		&			&$1$		&$1$				\\
			$r_0$  (\ans{arg1})  						&		&		&		&		&			&$2.9$		&$2.9$				\\
			\hline
			
			Coul-Buck~\cite{matsui:1996}							\\  		%	&		&		&		 &		 &		&		&		&		\\
			\hline
			$Z$ (e) 											&$1.4175$  	&$-0.945$		&		&		&		&				&		\\
			$A$ (eV) 													&		&		&$31574470$		&$28480$		&$6463.4$		&				&		\\
			$\rho$ (\ans{arg1}) 								&		&		&$0.068$		&$ 0.172$		&$0.276$		&		&		\\
			$C$ (eV \ans{arg1}$^6$)								&		&		&$14.07$		&$34.63$		&$85.22$				&		&		\\
			\hline
			
			B-M-H~\cite{bauchy:2014} 							\\  		%	&		&		&		 &		 &		&		&		&		\\
			\hline
			$Z$ (e) 											&$1.8$  	&$-1.2$		&		&		&		&				&		\\
			$A$ (eV) 													&		&		&$0.002896$		&$0.007489$		&$0.011984$		&				&		\\
			$\rho$ (\ans{arg1}) 								&		&		&$0.0680$		&$0.1640$		&$0.2630$		&		&		\\
			$\sigma$ (\ans{arg1})								&		&		&$1.5704$		&$2.6067$		&$3.6430$				&		&		\\
			$C$ (eV \ans{arg1}$^6$)								&		&		&$14.0305$		&$34.5272$		&$84.9671$				&		&		\\
			$D$ (eV \ans{arg1}$^8$)								&		&		&$0$		&$0$		&$0$				&		&		\\
			\hline
			
			CTI+EAM~\cite{streitz:1994, zhou:2004}  \\
			\hline     
			$\chi$ (eV) 											&$0$  	&$5.484763$		&		&		&		&				&		\\
			$J$ (eV) 											&$10.328655$  	&$14.035715$		&		&		&		&				&		\\
			$\gamma$ (\ans{arg1}$^{-1}$) 											&$0$  	&$0$		&		&		&		&				&		\\
			$\zeta$ (\ans{arg1}$^{-1}$) 											&$0.968438$  	&$2.143957$		&		&		&		&				&		\\
			$Z$ (e)																	&$0.763905$  	&$0$		&		&		&		&				&		\\
			$r_e$ (\ans{arg1})											&			 	&				&			&$2.511 075$		&$3.315 171$		&				&		\\
			$\alpha$ 																&			 	&				&			&$8.574 224$		&$5.716 137$		&				&		\\
			$\beta$																	&			 	&				&			&$4.669 743$		&$3.758 299$		&				&		\\
			$A$(eV)																		&			 	&				&		&$0.208 662$		&$0.263 795$		&				&		\\
			$B$(eV)																		&			 	&				&		&$0.678 293$		&$0.273 569$		&				&		\\
			$\kappa$																	&			 	&				&		&$0.355 898$		&$0.498 438$		&				&		\\
			$\lambda$																	&			 	&				&		&$1.014 487	$		&$0.560 282$		&				&		\\
			\hline 		
		\end{tabular}
		\caption{The potential parameters of four potentials. The parameters for \potCT{} are from the original source \cite{streitz:1994} and the parameters of EAM potential are from Ref.~\cite{zhou:2004}. }
	\end{table}
\end{center}

\newpage
\section*{Supplementary Material}
\setcounter{section}{0}
\setcounter{figure}{0}
\setcounter{equation}{0}
\renewcommand{\figurename}{Supplementary Figure}
\renewcommand\thefigure{S\arabic{figure}}

%\todo{Put in figure on calculation of melting temperature}

\subsubsection{Method of calculating melting temperature}
\begin{figure}[htbp!] %[htbp!]
	\centering
	\includegraphics[width=0.65\textwidth]{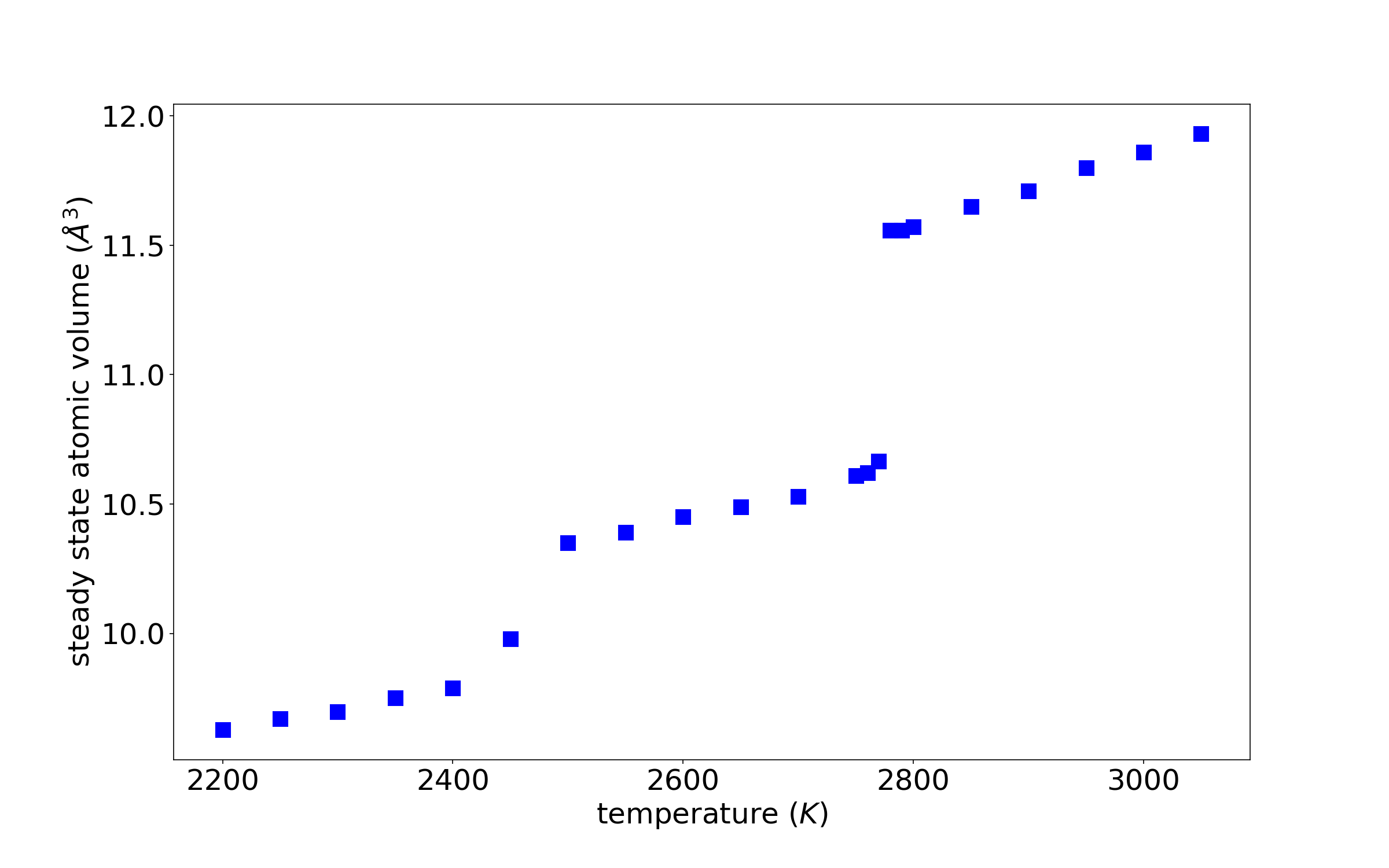} 
	\caption{The steady state volume per atom, obtained at various temperatures for \potV{} potential, is plotted with temperature. The window of temperature within which the volume jump occurs is considered to be the melting temperature of the potential. The melting temperature is $2760$~K  for \potV{}  potential associated with the biggest jump in volume per atom as shown in the figure.}
	\label{sfig:melting_temp_vash}
\end{figure}

An energetically minimized periodic alumina system is heated up to several temperatures for allowing volume expansion and equilibrated by NPT at the target temperatures for minimum of $100$~ps. The steady state volume per atom is plotted against temperatures for \potV{} in Fig.~\ref{sfig:melting_temp_vash} as an example.

\section{Information concerning the computation of material properties at 0~K}

  \subsection{Lattice constants, cohesive energy and vacancy formation energy}
 A 3D alumina system of $8640$ atoms $(3456 \; Al+5184 \; O)$ atoms, periodic in all directions are energetically minimized to calculate the energy per atom (cohesive energy) and lattice constants. The vacancy formation energy is calculated by taking a difference in energy between a minimized infinite pristine system and with a vacancy in it. 
 % The minimized system is minimized after individually removing Al and O atoms to calculate the vacancy formation energies for Al and O, respectively.
 
 \subsection{Surface energies}
 For surface energy calculations, infinite slabs with the desired crystallographic orientations in the free surface direction are minimised in energy. The value of the surface energy is then calculated as: 
 \begin{equation}
 	\gamma = \frac{E_{\rm slab}-E_{\rm coh}N}{2A},
 \end{equation}\label{eqn:surf_energy}
 where $E_{\rm slab} $ is the energy of the slab, $E_{\rm coh}$ is the cohesive energy, $A$ is the free surface area, and $N$ is the total number of atoms in the slab.

  \subsubsection{Identifying the appropriate slab thickness }
\begin{figure}[htbp!] %[htbp!]
	\centering
	\includegraphics[width=0.65\textwidth]{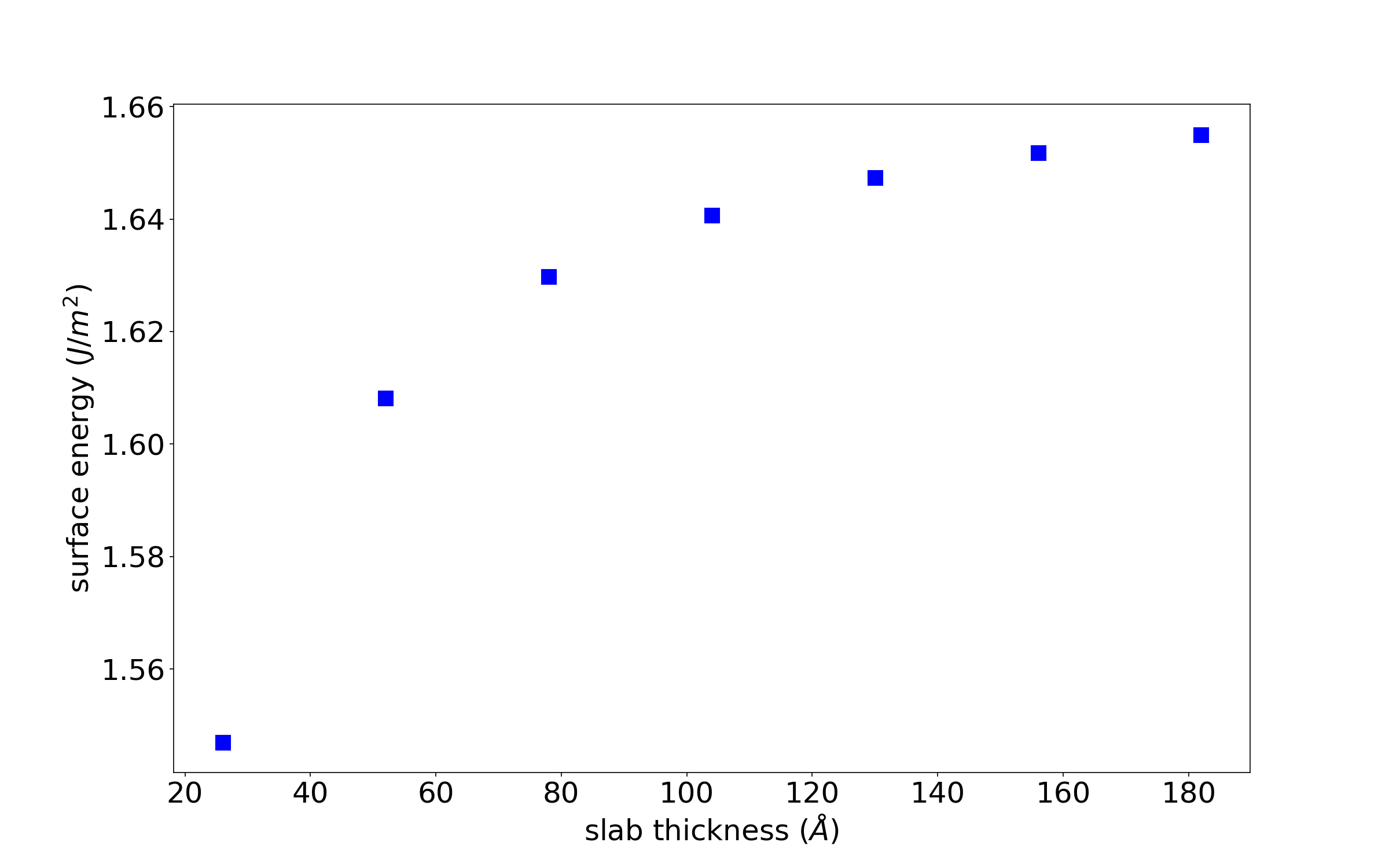} 
	\caption{Surface energy of $[0001]$ for slabs of different thicknesses calculated with \potV{} potential. }
	\label{sfig:gamma_thickness_vash}
\end{figure}

In order to minimise the influence of the thickness on the calculated surface energy, the same calculation is performed for several slab thicknesses. The calculated surface energy increases with the slab thickness, and it stabilises with respect to thickness beyond a certain value (see Fig.~\ref{sfig:gamma_thickness_vash}). Based on this study, around $60$~\ans\; is chosen for the minimum slab thickness to calculate surface energies.

 \subsection{Elastic constants}
 The system is deformed by applying a small strain in positive and negative Voigt strain directions. The elastic constants are calculated by taking the derivatives of the measured change in stress tensor with respect to strain.

 \subsection{Neck region}
\begin{figure}%[htbp!] %[htbp!]
 	\centering
 	\includegraphics[width=0.6\textwidth]{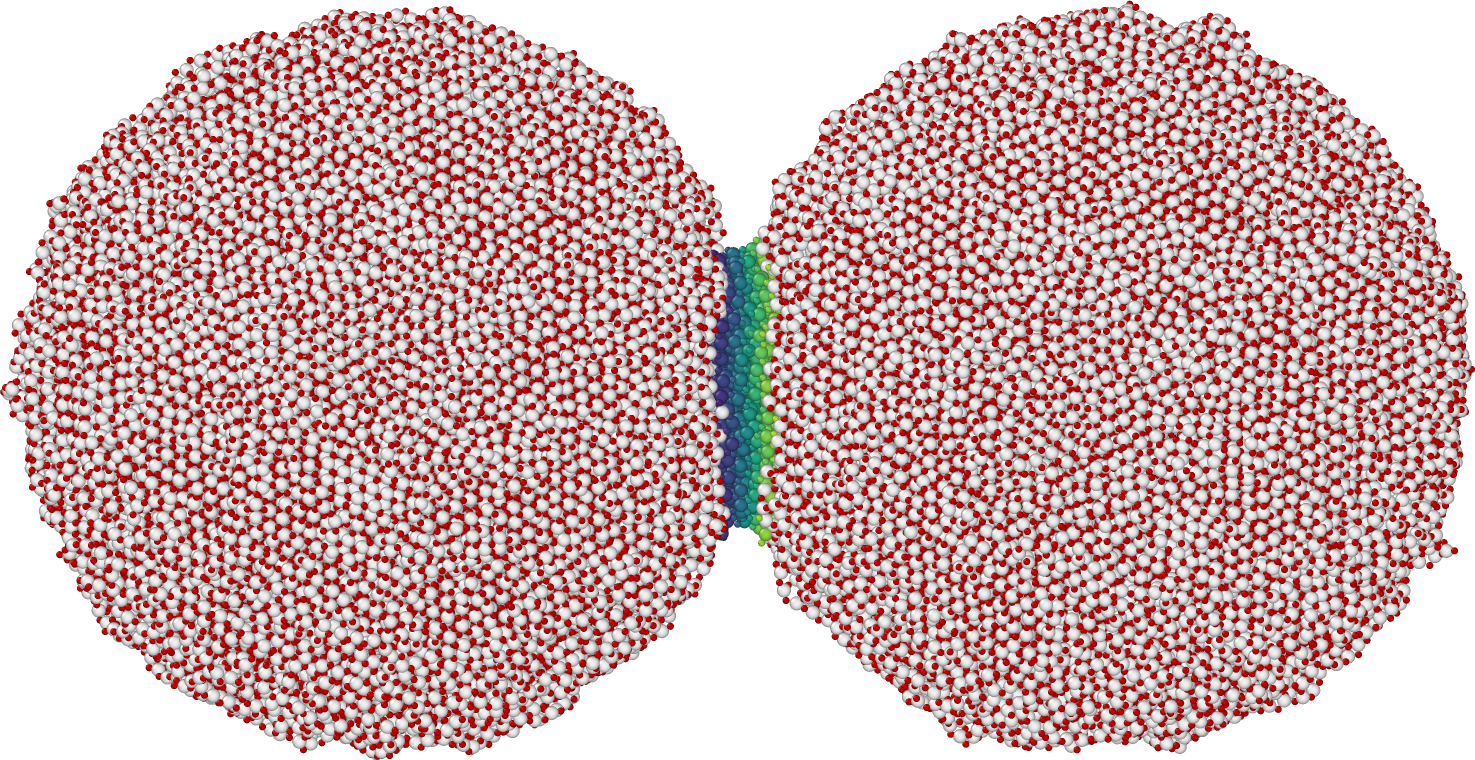} 
 	\caption{The neck region between two particles during sintering (highlighted with a color gradient). All other atoms are colored according to their chemical species: Al - gray, O - red. }
     \label{fig:neck_region}
 \end{figure}

In Fig.~\ref{fig:neck_region} the neck region of the sintering nanoparticles is highlighted by blue to green contour for \Rc{} particle and \potV. The details on defining a neck region is described in the main text.

 \subsection{Influence of rigid body motion of particles on MSD}
\begin{figure}[htbp!] 
 	\centering
 	\includegraphics[width=0.65\textwidth]{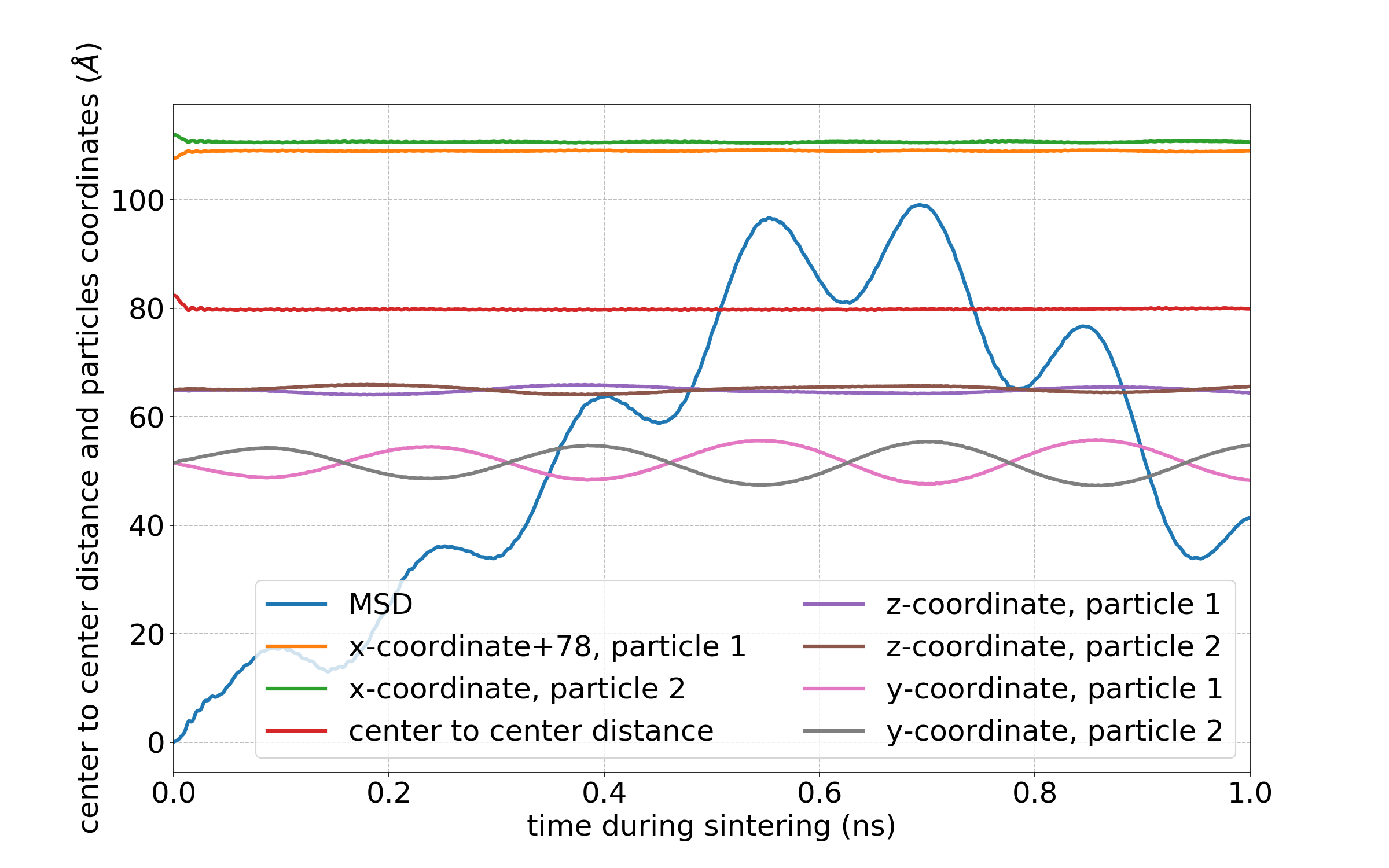} 
 	\caption{The rigid body motion pf particles contribute to the MSD calculation. The centre-to-centre distance between particles remains almost unchanged. The centre of masses of the particles do not change substantially in $x$ and $z$ direction. The particles oscillate with respect to their centre point in the $y$ direction resulting a local fluctuation in MSD.}\label{fig:rigid_body_msd}
 \end{figure}

 Fig.~\ref{fig:rigid_body_msd} shows the centre-to-centre distance and particle coordinates as a function of the sintering time.

\end{document}